%% file: main.tex
\documentclass[letterpaper,12pt]{article}
\usepackage[T1]{fontenc}
\usepackage[latin9]{inputenc}
\usepackage{bm}
\usepackage{amsmath}
\usepackage{amssymb}
\usepackage{amsfonts, mathrsfs}
\usepackage{eucal}
\usepackage{graphicx}
\usepackage[authoryear]{natbib}
\usepackage[unicode=true,
 bookmarks=false,
 breaklinks=false,pdfborder={0 0 1},backref=false,colorlinks=false]
 {hyperref}

\makeatletter

\providecommand{\tabularnewline}{\\}

\@ifundefined{date}{}{\date{}}
\allowdisplaybreaks

\usepackage{latexsym}
\usepackage{multirow}
\usepackage{url}% not crucial - just used below for the URL 
\usepackage{enumerate}
\usepackage{color, colortbl}
\usepackage[table]{xcolor}
\usepackage{mathtools}
\usepackage{array}
\usepackage{footnote}
\usepackage{tikz}
\usetikzlibrary{fit,positioning}
\usepackage{booktabs}
\usepackage{upgreek}

\definecolor{Gray}{gray}{0.9}
\def\bSig\mathbf{\Sigma}

\newcommand{\bg}{\bm{g}}
\newcommand{\blam}{\bm{\lambda}}
\newcommand{\tildelta}{\widetilde{\delta}}

\usepackage[colorinlistoftodos,textwidth=2.1cm]{todonotes}

 \newcommand{\indep}{\perp\!\!\!\!\perp}

\usepackage{colortbl}
\usepackage{subcaption}

\usepackage{dcolumn}
\newcolumntype{.}{D{.}{.}{-1}}
\newcolumntype{d}[1]{D{.}{.}{#1}}
\usepackage{theorem}
\theoremstyle{definition}
\newtheorem{assumption}{Assumption}

\newtheorem{theorem}{Theorem}

\def\T{{ \mathrm{\scriptscriptstyle T} }}

\usepackage{rotating}

\usepackage{arydshln}

\usepackage[compact]{titlesec}

\allowdisplaybreaks

\newcommand{\spacingset}[1]{\renewcommand{\baselinestretch}%
{#1}\small\normalsize}

\newcommand{\expit}{\text{expit}}

\newcommand{\V}{\mathcal{V}}
\newcommand{\calO}{\mathcal{O}}

\makeatother

\begin{document}
\title{\Large\textbf{Doubly robust estimators for generalizing treatment effects on survival outcomes from randomized controlled trials to a target population}}
\author
{\small\vspace{1ex} \textbf{Dasom Lee$^1$, Shu Yang$^{1,*}$ Xiaofei Wang$^{2}$}\\
\small \vspace{1ex}$^{1}$Department of Statistics, North Carolina State University, Raleigh, NC, U.S.A. \\
\small\vspace{1ex}$^{2}$Department of Biostatistics and Bioinformatics, Duke University, Durham, NC, U.S.A. \\
\small\vspace{1ex}$^{*}$\textit{email:} syang24@ncsu.edu}

\maketitle

\spacingset{1.3} 
\begin{abstract}
In the presence of heterogeneity between the randomized controlled trial (RCT) participants and the target population, evaluating the treatment effect solely based on the RCT often leads to biased quantification of the real-world treatment effect. To address the problem of lack of generalizability for the treatment effect estimated by the RCT sample, we leverage observational studies with large samples that are representative of the target population. This paper concerns evaluating treatment effects on survival outcomes for a target population and considers a broad class of estimands that are functionals of treatment-specific survival functions, including differences in survival probability and restricted mean survival times. Motivated by two intuitive but distinct approaches, i.e., imputation based on survival outcome regression and weighting based on inverse probability of sampling, censoring, and treatment assignment, we propose a semiparametric estimator through the guidance of the efficient influence function. The proposed estimator is doubly robust in the sense that it is consistent for the target population estimands if either the survival model or the weighting model is correctly specified, and is locally efficient when both are correct. In addition, as an alternative to parametric estimation, we employ the nonparametric method of sieves for flexible and robust estimation of the nuisance functions and show that the resulting estimator retains the root-$n$ consistency and efficiency, the so-called {\it rate-double robustness}. Simulation studies confirm the theoretical properties of the proposed estimator and show it outperforms competitors. We apply the proposed method to estimate the effect of adjuvant chemotherapy on survival in patients with early-stage resected non-small lung cancer.

\end{abstract}
%\newpage{}

\spacingset{1.5} 
\section{Introduction}
\label{sec:intro}
In clinical trials or biomedical studies, time-to-event or survival endpoints, such as the time from treatment initiation to death, have been commonly used to evaluate the treatment effect. When estimating treatment effect, randomized controlled trials (RCTs) are regarded as the gold standard since randomization reduces the effect of confounding variables.
However, RCTs often suffer from a lack of generalizability or external validity.
Specifically, due to restrictive inclusion and exclusion criteria for enrollment or additional concerns from patients and physicians, RCTs often don't recruit enough participants that represent the real-world patient population, resulting in the covariate distribution of the RCT sample being different from that of the target real-world population. In the presence of such heterogeneity, evaluating the treatment effect based solely on the RCT sample leads to biased quantification of the real-world treatment effect. As a complement of the RCT sample, observational studies have been widely used in comparative effectiveness research, as large samples that are representative of the target population can be studied at a relatively low cost.

Several recent works have proposed integrative methods to generalize findings from the RCT to the target population by leveraging observational studies \citep{cole2010generalizing, tipton2013improving, hartman2015sample, dahabreh2019generalizing, lee2021improving}. Most existing methods focus on directly modeling the probability of participating in the trial, i.e., the sampling score which is analogous to the treatment propensity score. 
A widely used approach involves inverse probability of sampling weighting \citep[IPSW;][]{cole2010generalizing, stuart2011use}, which can be used to estimate weight-adjusted survival curves \citep{cole2004adjusted, pan2008proportional}. 
However, these IPSW-based estimators are unstable under extreme sampling scores.
Alternatively, \citet{lee2021improving} proposed calibration weighting that enforces covariate balance between the RCT and observational study without explicitly modeling the sampling score.
Recently, \citet{colnet2020causal} provided a 
comprehensive review of various novel methods combining complementary features of RCTs and observational studies. However, most of these methods focus on continuous and binary outcomes, and generalization of the findings from RCTs for survival outcomes to the target population is less actively studied. 

In this paper, we consider a broad class of estimands defined as a functional of treatment-specific survival functions, including differences in survival probability at landmark times and restricted mean survival time (RMST). 
Various estimators can be constructed to adjust for the unrepresentativeness or selection bias of the RCT sample. 
One approach relies on fitting conditional survival outcome models and then averaging over the covariate distribution of the observational sample, similar to \citet{chen2001causal}. 
Another common approach is to use weighting \citep{cole2004adjusted, wei2008estimating} to adjust for the imbalance between the RCT sample and the observational sample. 
Instead of direct modeling of sampling score as in IPSW, one can consider the more stable approach that calibrates covariate distributions between the RCT and the observational sample \citep{lee2021improving}. 
Motivated by these two intuitive but distinct approaches, we propose improved estimators for survival outcomes under the guidance of the efficient influence function (EIF), which involve survival outcome regression and weighting based on inverse probability of treatment, censoring, and sampling, simultaneously.
The proposed estimator is doubly robust in the sense that it is consistent for the target population estimand if either the survival model or the weighting model is correctly specified, and is locally efficient when both are correct. In addition, to cope with possible misspecification of nuisance functions, we consider the method of sieves \citep{chen2007large}, which adds great flexibility and robustness to the proposed estimators, meanwhile retaining the root-$n$ consistency.

The remainder of the paper is organized as follows. In Section \ref{sec:basic setup}, we formalize the basic causal inference framework for survival outcomes. In Section \ref{sec:method}, we introduce two direct estimators based on identification formulas, and in Section \ref{sec:improved estimators}, we propose improved estimators and describe the corresponding asymptotic properties. The finite sample performance of the proposed estimators is assessed via simulation studies in Section \ref{sec:simulation}. Applying the proposed estimators, we analyze the effect of adjuvant chemotherapy on the survival of lung cancer patients with data from an RCT and an observational study in Section \ref{sec:application}. Section \ref{sec:discussion} presents the discussion and concluding remarks. All proofs are provided in the Appendix.

\section{Estimands, Observed Data, and Assumptions}
\label{sec:basic setup}
Suppose we are interested in comparing the effectiveness of two treatments. Let $A$ be the binary treatment assignment, A $\in \{0, 1 \}$. Following the potential outcomes framework \citep{rubin1974estimating, rubin1986comment}, let $T^a$ be the potential survival time if a subject received the treatment $A = a$, and $S_a(t)$ and $\lambda_a(t)$ be the corresponding survival and hazard functions, i.e., $S_a(t) = P(T^a \ge t)$ and $\lambda_a(t) = \lim_{h \to 0} h^{-1} P(t \le T^a \le t + h) / P(T^a \ge t)$. 
Under the proportional hazards assumption, a widely used measure to characterize the treatment effect is hazard ratio (HR), i.e., $\lambda_1(t)/\lambda_0(t)$ being a constant. However, the interpretation of HRs is challenging especially when the proportionality assumption is violated \citep{hernan2010hazards, trinquart2016comparison}.

Alternatively, we define the average treatment effect (ATE) measure $\theta_{\tau}$ as a function of treatment-specific survival curves, $\theta_{\tau} = \Psi_{\tau}\left(S_1(t), S_0(t)\right)$ where $\tau$ is a pre-specified constant.
%for a properly chosen $\Psi_{\tau}(\cdot)$ that depend on a pre-specified constant $\tau$ .
%$\Psi_{\tau}\left(S_1(t), S_0(t)\right) \simeq \int_0^\tau \psi_1(t)S_1(t) \mathrm{d}t + \int_0^\tau \psi_0(t)S_0(t) \mathrm{d}t$ %for some $\psi_a(t), a \in \{0, 1\}$. The symbol $\simeq$ represents equal to or asymptotically equal to, and $\tau$ is a pre-specified constant. 
%This definition enables constructing the semiparametric efficient estimator of $\theta_{\tau}$ based on the estimated treatment-specific survival curves.
This formulation of the ATE includes a broad class of estimands that are favored in survival analysis \citep{yang2020smim}.  
For example, $\theta_{\tau} = S_1(\tau) - S_0(\tau)$ is a simple survival difference at a fixed time $\tau$, %with $\psi_1(t) = -\psi_0(t) = I(t = \tau)$,
and $\theta_{\tau} = \int_0^{\tau} \{S_1(t) - S_0(t)\} \mathrm{d}t$ is the restricted mean survival time (RMST) difference up to $\tau$. %, with $\psi_1(t) = -\psi_0(t) = 1$.  
The ratio of restricted mean time loss (RMTL) and the difference of the median survival can also be represented with the appropriate choice of $\Psi_{\tau}(\cdot)$.

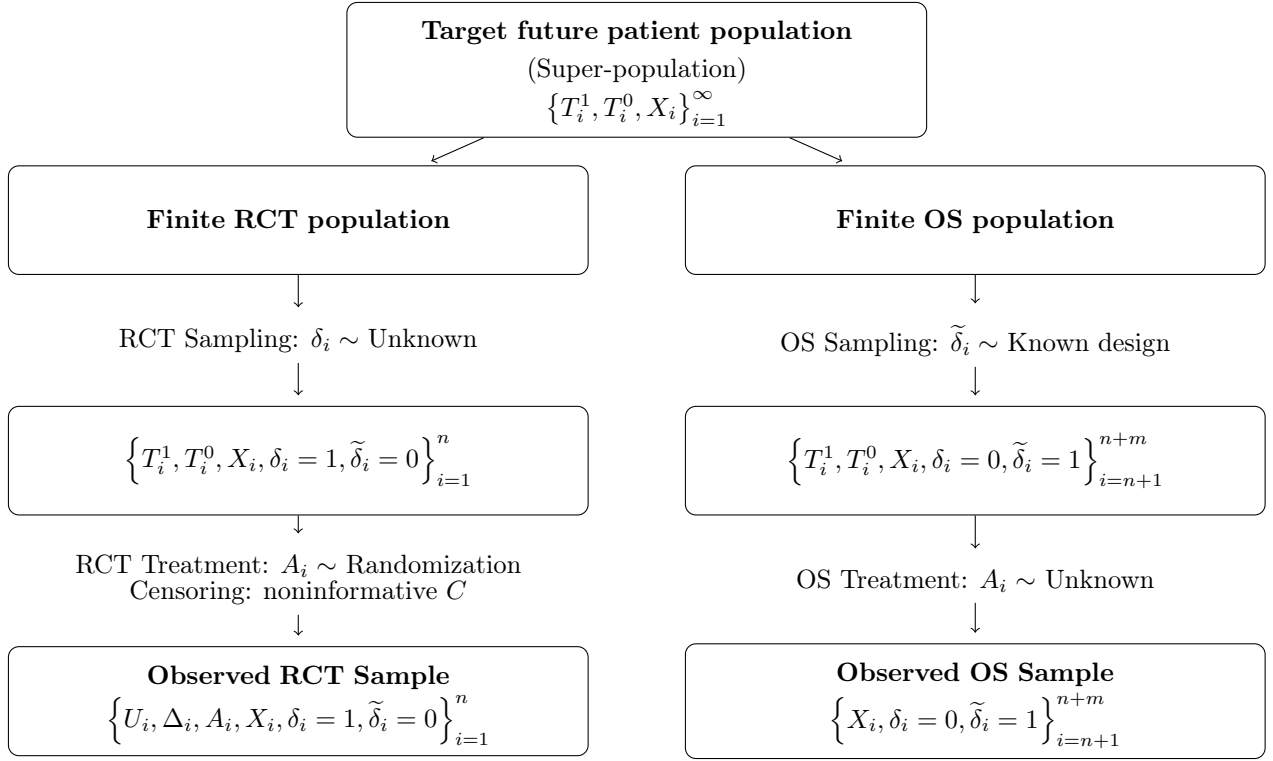
\begin{figure}[t]
\begin{centering}
\input{demo.tex}
\par\end{centering}
\caption{Illustration of the data structure of the RCT sample and the observational study (OS) sample within the target super-population framework.}
\label{fig:data structure}
\end{figure}

Under the Stable Unit Treatment Value Assumption, the survival time $T$ is the realization of the potential outcomes, i.e., $T = T^1A + T^0(1-A)$. Let $C$ be the censoring time. In the presence of right censoring, the survival time $T$ is not observed for all subjects; instead, we observe $U = T \wedge C$ and $\Delta = I(T \le C)$ where $\wedge$ represents the minimum of two values, and $I(\cdot)$ is an indicator function. 
Let $X$ be a $p$-dimensional vector of pre-treatment covariates. 
Also, let $\delta$ denotes the binary indicator of RCT participation, and let $\widetilde{\delta}$ denotes the binary indicator of observational study participation. We consider a super-population framework assuming that an RCT sample of size $n$ and an observational sample of size $m$ are sampled from the target population. From the RCT sample, we observe $\{U_i, \Delta_i, A_i, X_i, \delta_i = 1, \widetilde{\delta}_i = 0\}$ from $i = 1,...,n$ independent and identically distributed subjects. For the observational sample, it is common that only the covariates information is available, i.e., $\{X_i, \delta_i = 0, \widetilde{\delta}_i = 1\}$ from $i = n+1 ,...,n+m$ independent and identically distributed subjects. 
The sampling mechanism and data structure are illustrated in Figure \ref{fig:data structure}. We assume independence between the RCT and the observational sample, which holds if two separate studies are conducted by independent researchers, the target patient population is sufficiently large, or the patients are enrolled in two separate time periods.

%In survival analysis, identification of the causal estimand is often plagued by missing potential outcomes and censoring.
Let $S_a(t \mid X) = S(t \mid X, A = a, \delta = 1)$ be the treatment-specific conditional survival curves for $a, \delta \in \{0, 1\}$.
Also, define the treatment propensity score $\pi_{A}(X) = P(A = 1 \mid X, \delta = 1)$ and the sampling score $\pi_{\delta}(X) = P(\delta = 1 \mid X)$.
In order to identify the ATE from the observed data, we make the following assumptions:

\begin{assumption}[Ignorability and positivity of trial treatment assignment]\label{assump:treatment ignorability} \textcolor{white}{1}\\
(i) $\{T^0, T^1\} \indep A\mid(X,\delta=1)$; and (ii) $0<\pi_{A}(X)<1$ with probability $1$.
\end{assumption}

\begin{assumption}[Conditional survival exchangeablity and positivity of trial participation]\label{assump:trial ignorability} 
(i) $S_a(t \mid X, \delta = 1) = S_a(t \mid X), a \in \{0, 1 \}$; and (ii) $\pi_{\delta}(X)>0$  with probability $1$.
\end{assumption}

\begin{assumption}[Noninformative censoring conditional on covariates and treatment]\label{assump:noninformative censoring} \textcolor{white}{1}\\
$\{T^1, T^0\} \indep C \mid (X, A, \delta = 1)$, which also implies $T \indep C \mid (X, A, \delta = 1)$.
\end{assumption}
Assumptions \ref{assump:treatment ignorability}--\ref{assump:noninformative censoring} are not testable in general and their plausibility should be justified based on subject matter knowledge in practice. Assumption \ref{assump:treatment ignorability} holds in the RCT by default. Assumption \ref{assump:trial ignorability} (i) is plausible if all information related to the trial participation and the outcome is captured in the data at hand. This assumption is weaker than the ignorablility of the trial participation assumption, i.e.,  $\{T^0, T^1\} \indep\delta\mid X$. The relationship between Assumption \ref{assump:trial ignorability} (i) and its stronger version is analogous to that described in \citet{dahabreh2019generalizing} in the context of continuous and binary outcomes.
Assumption \ref{assump:trial ignorability} (ii) implies that the absence of patient characteristics that prevent from participating in the trial \citep{lee2021improving}.
Assumption \ref{assump:noninformative censoring} is commonly made in survival analysis \citep{chen2001causal, zhang2012double, zhang2019estimating}. This assumption is weaker than the conditional independence assumption of the censoring and survival time given only on the treatment \citep{zhang2012contrasting}.

The covariate distribution of the RCT sample $f(X \mid \delta = 1)$ may not be representative of that of the target population $f(X)$ due to restrictive trial enrollment criteria, but the covariate distribution of the observational sample $f(X \mid \tildelta = 1)$ is often representative of $f(X)$ due to the real-world data collection mechanism. In particular, if the observational sample is a simple random sample of the target population, then $f(X \mid \tildelta = 1)=f(X)$. More generally, the observational sample can be selected under complex sampling designs. To accommodate such
scenarios, we can define $d$ as the known design weight for the observational sample. 

\begin{assumption}[The known design weight for the observational sample]\label{assump:OSXdis}\textcolor{white}{1}\\
The observational sample design weight $d = 1/P(\widetilde{\delta} = 1 \mid X)$ is known.
\end{assumption}
Assumption \ref{assump:OSXdis} is commonly assumed in the survey sampling literature. Based on Assumption \ref{assump:OSXdis}, the design-weighted observational sample is representative of the target population.  In an observational study with simple random sampling, $d = N/m$, where $N$ is the target population size.

Under the above assumptions, the ATE $\theta_{\tau}$, or $S_a(t), \  a \in \{0, 1\}$ sufficiently, are identified based on the observed data. 
We consider two identification formulas. Let $Y(t) = I(U \ge t)$, and define the conditional censoring model $S^C(t\mid X, A) = P(C > t \mid X, A, \delta = 1)$. 
One identification formula is based on the conditional survival curves, i.e.,
\begin{align}
    S_a(t) = \mathbb{E} \left\{\widetilde{\delta}d S_a(t \mid X) \right\}, ~
    S_a(t \mid X) = \mathbb{E}\{I(T \ge t) \mid X, A =a, \delta = 1 \},
    \label{eq:identification-outcome}
\end{align}
which can be called the OS-design-weighted G-computation formula. Note that if the population covariate distribution is available, $\mathbb{E}\{ S_a(t,X)\}$ is the G-computation formula for $S_a(t)$.
The other identification formula is based on the inverse probability weighting (IPW) approach for the marginal survival curves, i.e.,
\begin{align}
\label{eq:identification-weighting}
S_a(t) = \mathbb{E}\left[\frac{\delta}{\pi_{\delta}(X)}\frac{I(A=a)}{\pi_{A}(X)^a\{1-\pi_A(X)\}^{1-a} }\frac{Y(t)}{S^C(t\mid X, A)}\right],
\end{align}
for $a \in \{0,1\}$. The two identification formulas in \eqref{eq:identification-outcome} and \eqref{eq:identification-weighting} motivate the estimators in the following section, depending on different components of the observed data likelihood.

%\sy{Explicitly saying that there are two identification formulas one based on the outcome model and the other based on the weighting, which motivate the two estimators and the proposed one using the EIF.}

%\[\tau = \int_0^L \left \{ S_1(t) - S_0(t) \right \} dt = \int_0^L \left [ \mathbb{E} \left\{\widetilde{\delta} S(t \mid X, A = 1, \delta = 1) \right\}- \mathbb{E} \left\{\widetilde{\delta}S(t \mid X, A = 0, \delta = 1) \right\} \right ] dt. \] 

%Let $T$ be the time to event and and $C$ be the time to censoring. 
%Due to the censoring, instead of $T$, we observe $\{U, \Delta\}$, where $U = T \wedge C$ and $\Delta = I(T \le C)$, where $I(\cdot)$ is the indicator function. 

%Let A be the binary treatment assignment, A $\in \{0, 1 \}$, $T^a$ be the potential survival time if a subject received treatment $a$. We define the average causal treatment effect as $\tau_L = \int_0^L \{P(T^1 \ge t) - P(T^0 \ge t)\} dt$, the difference in the mean of the restricted lifetime up to a specific time $L$. 

% Conditional survival time $S_a(t \mid X = x) = P(T \ge t \mid A = a, X = x)$.
%Estimate $S_a(t)$ by $N^{-1}\sum_{i=1}^N \delta_i \widehat{S}_a(t \mid X = x_i)$ where $\widehat{S}_a(t \mid X = x_i)$ estimates $P(T \ge t \mid A = a, X = X_i)$

\section{Two Direct Estimators based on Identification Formulas}
\label{sec:method}

\subsection{Outcome Regression}
\label{sec:OR}
Based on the identification formula \eqref{eq:identification-outcome}, the treatment-specific conditional survival curve can be modeled and fitted based on the observed data, e.g., using the widely used Cox regression model for the survival outcome.
A treatment-specific conditional hazard function at time $t$ given covariate $X_i$ is defined as
\begin{equation}
    \lambda_{ai}(t) \equiv \lambda_a(t \mid X_i) = \lambda_{a0}(t)\exp(\beta_a^{\T}X_i), \label{eq:cox model}
\end{equation}
where $\lambda_{a0}(t)$ is a treatment-specific baseline hazard function, for $a \in \{0, 1 \}, i = 1,...,n$. Following standard survival analysis techniques, $\beta_a$ can be estimated as a solution to the partial likelihood score equation, and the baseline cumulative hazard $\Lambda_{a0}(t) \equiv \int_0^t \lambda_{a0}(u)$ can be estimated by the \citet{breslow1974covariance} estimator,
$$\hat{\Lambda}_{a0}(t) = \int_0^t \frac{\sum_{i=1}^N \delta_i A_{ai} \mathrm{d}N_i(u)} {\sum_{i=1}^N \delta_i A_{ai} \exp(\widehat{\beta}_a^{\T}X_i)Y_i(u)}, $$
where $A_{ai} = I(A_i = a)$, $N_i(u) = I(U_i \le u, \Delta_i = 1)$, and $Y_i(u) = I(U_i \ge u)$. The survival model in \eqref{eq:cox model} does not imply that the marginal HR of the potential survival outcomes under $a=1$ and $a=0$, i.e.,  $\lambda_{10}(t)/\lambda_{00}(t)$, is a constant and thus is not restrictive. Other survival models can be considered, including the additive hazards model \citep{lin1995semiparametric, aalen1989linear}.
%\sy{add a ref}.
%implies that the proportional hazards assumption is not restrictive, as $\lambda_a(t)$ and $\beta_a$ can vary by treatment.

\citet{chen2001causal} proposed a method that accounts for imbalances between treatment groups by first estimating the conditional treatment effect given $X$ and then averaging the effect over $X$ across both treatment groups. 
A similar approach can be applied to balance the covariate distribution between the RCT sample and the observational sample by first estimating the treatment-specific survival curve conditional on $f(X \mid \delta = 1)$ under model \eqref{eq:cox model}, i.e., $\widehat{S}_a(t \mid X_i) = \exp \left \{- \widehat{\Lambda}_{ai}(t)\right\} =  \exp \left \{- \widehat{\Lambda}_{a0}(t) \exp(\widehat{\beta}_a^{T}X_i) \right \}$, 
and then applying the design-weighted averaging over $f(X \mid \tildelta = 1)$. 
%Specifically, under model \eqref{eq:cox model}, we first estimate the conditional survival curve by $\widehat{S}_a(t \mid X_i) = \exp \left \{- \widehat{\Lambda}_{ai}(t)\right\} =  \exp \left \{- \widehat{\Lambda}_{a0}(t) \exp(\widehat{\beta}_a^{T}X_i) \right \}$, and then average over the covariates of the observational sample, 
The resulting outcome regression (OR) estimator of the marginal treatment-effect survival curve is
\begin{align}
    \widehat{S}^{\text{OR}}_a(t) = & \left(\sum_{i=1}^{N} \tildelta_i d_i\right)^{-1} \sum_{i=1}^N \tildelta_i d_i e^{-\widehat{\Lambda}_{ai}(t)},
    %= & m^{-1} \sum_{i=1}^N \tildelta_i e^{-\widehat{\Lambda}_{ai}(t)} \mbox{   in the simple random sampling},
    \label{eq:OR estimator}
\end{align}
for $a \in \{0, 1\}$, and the corresponding
%Consequently, the OR estimator of the 
ATE estimator is defined as $\widehat{\theta}^{OR}_{\tau} = \Psi_{\tau}\left(\widehat{S}^{\text{OR}}_1(t), \widehat{S}^{\text{OR}}_0(t)\right)$. The OR estimator is consistent when the survival model \eqref{eq:cox model} is correctly specified.

\subsection{Inverse Probability and Calibration Weighting}
\label{sec:CW}
We can construct the estimator of the marginal  treatment-specific survival curve based on the identification formula in  \eqref{eq:identification-weighting} that involves sampling score, treatment propensity score, and censoring probability. This approach can be viewed as a combination of IPSW, inverse probability of treatment weighting (IPTW), and inverse probability of censoring weighting (IPCW). The weighting estimator requires positing models for the three probabilities and estimating them.

First, we consider the estimation of the sampling score. As the covariate distribution of the RCT sample is different from that of the target population in general, the estimated ATE based only on the RCT sample can be biased.
 A widely used approach to account for this selection bias is IPSW, i.e., weighting the RCT  sample by the inverse probability of trial participation over that of observational study participation, to adjust for differences in covariate distribution between the trial sample and the population. 
 Specifically, $\pi_{\delta}(X)$ can be modeled as $\pi_{\delta}(X) = \{{\omega}_{\text{IPSW}}(X)\}^{-1}$ where 
 ${\omega}_{\text{IPSW}}(X) =  P(\tildelta = 1 \mid \delta + \tildelta = 1, X)/P(\delta = 1 \mid \delta + \tildelta = 1, X)$. One can plug in $\{\widehat{\omega}_{\text{IPSW}}(X)\}^{-1}$ for $\pi_{\delta}(X)$ in \eqref{eq:identification-weighting} using the common logistic regression model.
However, the IPSW method requires the sampling score model to be correctly specified; it also could be highly unstable if $P(\delta = 1 \mid \delta + \tildelta = 1, X)$ is close to zero for some $X$.

Instead of direct estimating the sampling scores, \citet{lee2021improving} proposed the calibration weighting approach to reduce the selection bias in the trial-based estimator, which is analogous to the entropy balancing method by \citet{hainmueller2012entropy} and more stable than the IPSW method. 
The basic idea is that subjects in the RCT sample are calibrated to the observational sample, so that after calibration, the covariate distribution of the RCT sample empirically matches that of the target population. The calibration weighting approach is based on the idea that for any $\bg(X)$, the following identity hold,
\begin{equation}
E \left\{ \frac{\delta}{\pi_{\delta}(X)} \bg(X) \right\} = E \left \{\widetilde{\delta} d \bg(X) \right \} = E \{\bg(X)\}, \label{eq:identity}
\end{equation}
where $\bg(X)$ is a function of $X$ to be calibrated, e.g., the moments or any nonlinear transformations.

The calibration weights $\omega_i$ are obtained by solving the optimization problem
\begin{align}
    \min_{\mathcal{W}}&\sum_{i=1}^n \omega_i\log \omega_i \label{eq:optimization},\\
    \mbox{subject to } & \omega_i \ge 0, ~\forall i,  \sum_{i=1}^n \omega_i = 1, \mbox{ and } \sum_{i=1}^N\delta_i\omega_i \bg(X_i) = \widetilde{\bm{g}}, \notag
\end{align}
   where $\mathcal{W} = \{w_i: \delta_i = 1\}$.
   The last constraint is the empirical representation of \eqref{eq:identity}, where $\widetilde{\bm{g}} = \sum_{i=1}^N\widetilde{\delta}_id_i\bg(X_i)/\sum_{i=1}^N\widetilde{\delta}_id_i$ is a consistent estimator of $\mathbb{E}\{\bg(X)\}$ from the observational sample. Minimizing the negative entropy function of the calibration weights in \eqref{eq:optimization} enforces the weights to be as close to one another as possible, which reduces the variability due to heterogeneous weights. 
    Using the Lagrange multiplier $\blam$, the objective function of this convex optimization problem becomes 
$L(\blam,\mathcal{W})=\sum_{i=1}^{n}\omega_{i}\log \omega_{i}-\blam^{\top}\left\{ \sum_{i=1}^{n}\omega_{i}\bg(X_{i})-\widetilde{\bg}\right\}$.
The estimated calibration weights are $\widehat{\omega}_{i}=\omega(X_{i};\widehat{\blam})=\exp\{ \widehat{\blam}^{\top}\bg(X_{i})\} / [\sum_{i=1}^{n}\exp\{ \widehat{\blam}^{\top} \bg(X_{i})\} ]$,
where $\widehat{\blam}$ solves
\begin{align}
    U(\blam)=\sum_{i=1}^{n}\exp\left\{ \blam^{\top}\bg(X_{i})\right\} \left\{ \bg(X_{i})-\widetilde{\bg}\right\} =0.\label{eq:dualproblem}
\end{align}
Under the loglinear sampling score model, the calibration weights from the objective function \eqref{eq:optimization} have the same functional form as inverse probability of sampling score weights asymptotically, resulting in the direct correspondence between the calibration weight and the  sampling score in that $\widehat{\omega}_{i}- \{N\widehat{\pi}_{\delta}(X_i)\}^{-1} \overset{p}{\to}0$, as $n\rightarrow \infty$ \citep{lee2021improving}.
%\citet{lee2021improving} proved that if the sampling score follows a loglinear model, then the calibration weight corresponds to the estimated sampling score in that $\widehat{\omega}_{i}- \{N\widehat{\pi}_{\delta}(X_i)\}^{-1} \overset{p}{\to}0$, as $n\rightarrow \infty$. 
%\sy{check to if you have defined $N$ somewhere.} - Defined when introduced design weight (pg 4)
Following that, we posit the loglinear sampling score model,
\begin{equation}
    \pi_{\delta}(X)=\exp\{\eta_{0}^{T}\bg(X)\}, \mbox{ for some } \eta_{0}.\label{eq:sampling model}
\end{equation}
\citet{lee2021improving} showed that $\widehat{\blam}$ is equivalent to $-\widehat{\eta}$, where $\widehat{\pi}_{\delta}(X)=\pi_{\delta}(X ; \widehat{\eta}) = \exp\{\widehat{\eta}^{T}\bg(X)\}$. Accordingly, in the rest of the paper, we represent the calibration weights using $\widehat\eta$, i.e., $\widehat{\omega}_{i}=\omega(X_{i};\widehat{\eta})=\exp\{ -\widehat{\eta}^{\top}\bg(X_{i})\} / [\sum_{i=1}^{n}\exp\{- \widehat{\eta}^{\top} \bg(X_{i})\}]$.
If one considers a logistic sampling score model instead, then other objective functions can be used, such as $\sum_{i=1}^n(\omega_i - 1)\log(\omega_i - 1)$, that corresponds to the weights with the same functional form as the inverse to a logistic probability of sampling \citep{zhao2019covariate, josey2020framework}. However, the loglinear regression model in \eqref{eq:sampling model} is close to the logistic regression model when the proportion of the RCT sample in the target population is small. 
%\sy{here need to mention that if one considers a logistic sampling score model, then one can change the objective function to .... But here we consider (6) because (some advantages of EB as mentioned in Lee's paper).}

With respect to treatment assignment, $\pi_A(X)$ is generally known for RCTs. However, several authors suggested estimating the treatment propensity score for the RCTs in order to increase the efficiency and account for the chance of imbalance of prognostic variables 
\citep[e.g.,][]{williamson2014variance, colantuoni2015leveraging}. We choose a logistic regression model for the treatment propensity score,
\begin{equation}
    \pi_{A}(X)=\left[1 + \exp\{-\rho_{0}^{T}\bg(X)\} \right]^{-1}, \mbox{ for some } \rho_{0}\label{eq:propensity model},
\end{equation}
and define $\widehat{\pi}_{ai} = A_i\pi_{A}\left(X_i ; \widehat{\rho}\right) + (1-A_i)\left\{1-\pi_{A}\left(X_i ; \widehat{\rho}\right)\right\}$. 
Estimating the propensity scores also broadens the scope of the current paper to allow the generalization of the observational study. Even though this paper focuses on generalizing the trial findings where we only require a two-way balancing between the RCT and observational study, a three-way balancing approach between the treated, the controlled, and the observational sample (e.g., \citep{chan2016globally}) can be useful to generalize the findings from the observational study to its larger population with the estimation of ${\pi}_{A}$.

%results to its larger population which is analogous to a three-way balancing approach between the treated, the controlled, and the observational sample \citep{chan2016globally}.
Moreover, in order to account for right censoring $C$, we posit Cox proportional hazards model with conditional hazard 
\begin{align}
    \lambda^C(t \mid X, A = a) = \lambda^C_{a0}(t)\exp(\gamma_a^TX), \mbox{for } a \in \{0, 1\}\label{eq:censoring model},
\end{align}
where the standard techniques as for the survival model in \eqref{eq:cox model} can be used to estimate $\gamma_a$ and $\Lambda^C_{a0}(t) \equiv \int_0^t \lambda^C_{a0}(u)\rm{d}u$. %Under Assumption \ref{assump:noninformative censoring}, the expectation of $Y(t)/S_C(t \mid X, A)$ leads to the conditional expectation of the survival time given $(X, A)$ in the RCT. 

Combining $\widehat{\omega}_i, \widehat{\pi}_{ai}$, and $\widehat{\Lambda}_{ai}^C(t) = \widehat{\Lambda}^C_{a0}(t) \exp(\widehat{\gamma}_a^{T}X_i)$ estimated under the working models in \eqref{eq:sampling model}, \eqref{eq:propensity model}, and \eqref{eq:censoring model}, respectively,
we define the CW estimator of the marginal treatment-specific survival curve as
\begin{equation}
    \widehat{S}_a^{CW}(t) = \sum_{i=1}^N\delta_i\widehat{\omega}_i\frac{A_{ai}}{\widehat{\pi}_{ai}}e^{\widehat{\Lambda}_{ai}^C(t)}Y_{i}(t) 
\label{eq:cw estimator},
\end{equation} 
where $A_{ai} = I(A_i = a)$. The corresponding ATE estimator is $\widehat{\theta}^{CW}_{\tau} = \Psi_{\tau}\left(\widehat{S}^{CW}_1(t), \widehat{S}^{CW}_0(t)\right)$.

\section{Improved Estimators}
\label{sec:improved estimators}

\subsection{Efficient Influence Function}
\label{sec:EIF}

Let $\mathcal{O}=(X,A,U, \Delta,\delta,\tildelta)$ be one copy of the vector of \textit{observed variables}. %, and let the likelihood function be
%\[f(\mathcal{O}) = \left\{ f(X)\pi_{\delta}(X)f(A \mid X, \delta = 1)f(U, \Delta \mid X, A,  \delta = 1) \right\}^{\delta} \left\{ f(X) \right \}^{\tildelta}.\]
The OR estimator specified in \eqref{eq:OR estimator} and the CW estimator specified in \eqref{eq:cw estimator} use different components of the likelihood function $f(\mathcal{O})$. 
Specifically, the OR estimator is based on modeling $  S_a(t \mid X)$ for $a\in\{0,1\} $, and the CW estimator is based on modeling $\pi_\delta(X)$, $\pi_a(X)$, and $S^C(t\mid A=1,X)$ for $a\in\{0,1\}$. 
These estimators are singly robust in that they are consistent only under the correct survival outcome regression model or the correct weighting models.
A vast number of estimators can be constructed by combining these two estimators. In general, the question becomes how to obtain the most efficient estimator. 
Our approach is to derive the EIF \citep{tsiatis2006semiparametric} of $\theta_\tau$ to construct semiparametrically efficient estimators. Such estimators also gain robustness to model misspecification as we show later. 

We consider a class of influence functions of regular asymptotically linear (RAL) estimators of the treatment-specific survival curve $S_a(t)$. 
Define $A_a = I(A = a)$ and $\pi_{a}(X) = a\pi_{A}(X) + (1-a)\{1-\pi_{A}(X)\}$.
Following \citet{tsiatis2006semiparametric}, the class of observed data influence functions includes
\begin{align}
    \varphi_a(t;\mathcal{O}) = & \frac{\delta}{\pi_{\delta}(X)}\frac{A_a}{\pi_{a}(X)}\frac{Y(t)}{S^C(t\mid A=a, X)}%^{\text{the RAL estimator}}
    - S_a(t)\notag\\
            & + \overbrace{\frac{\delta}{\pi_{\delta}(X)}\{A - \pi_A(X)\}g_{a1}(X)}^{\text{an arbitrary score of } A \mid (X, \delta = 1)} + \overbrace{\left\{ \frac{\delta}{\pi_{\delta}(X)} - \tildelta d \right\}g_{a2}(X)}^{\text{an arbitrary score of } \delta \mid X} \notag\\
            & + \overbrace{\frac{\delta}{\pi_{\delta}(X)}\frac{A_a}{\pi_{a}(X)}\int_0^t \frac{\mathrm{d}M^C_a(u)}{S^C(u\mid A=a, X)}g_{a3}(u\mid A=a, X)}^{\text{an arbitrary score of } U, \Delta = 0 \mid (X, A, \delta = 1)}
            \label{eq:class of influence}
\end{align}
for arbitrary functions $g_{a1}(\cdot), g_{a2}(\cdot)$, and $g_{a3}(\cdot)$, where $M^C_a(u) = N_a^C(u) - \int_0^u\Lambda_a^C(s) ds$ is a Martingale with $N_a^C = A_aI(U \le u, \Delta = 0)$, and $\Lambda_a^C(s)$ is a cumulative hazard for censoring for $a \in \{0, 1\}$. The last three terms in \eqref{eq:class of influence} are mean-zero functions. 
According to the semiparametric theory \citep{tsiatis2006semiparametric}, the EIF is the influence function in \eqref{eq:class of influence} with the smallest variance. The EIF can be derived by projecting %the influence function of a simple RAL estimator, i.e., 
the first term of \eqref{eq:class of influence} onto the orthogonal complement of the tangent space spanned by the scores of the nuisance functions, i.e., the last three terms. 
The RAL estimator with the EIF is a semiparametrically  efficient estimator.
The following theorem gives the EIF in the class of influence functions \eqref{eq:class of influence}, with the proof given in the  Appendix \ref{s:proof EIF}.

\begin{theorem}\label{thm:EIF}
Under Assumptions \ref{assump:treatment ignorability}--\ref{assump:OSXdis},
the EIF for the treatment-specific survival curve is
\begin{align}
    \varphi_{a}^{\text{eff}}(t;\mathcal{O}) = & \frac{\delta}{\pi_{\delta}(X)}\frac{A_a}{\pi_{a}(X)}\frac{Y(t)}{S^C(t\mid A=a, X)} - S_a(t) \notag\\
            & - \frac{\delta}{\pi_{\delta}(X)}\left\{\frac{A - \pi_A(X)}{\pi_A(X)}\right\}\mathbb{E}\{I(T \ge t) \mid X, A =a, \delta = 1 \} \notag\\
            & - \left\{ \frac{\delta}{\pi_{\delta}(X)} - \tildelta d \right\}\mathbb{E}\{I(T \ge t) \mid X, A =a, \delta = 1 \} \notag\\
            & + \frac{\delta}{\pi_{\delta}(X)}\frac{A_a}{\pi_{a}(X)}\int_0^t \frac{\mathrm{d}M^C_a(u)}{S^C(u\mid A=a, X)}\mathbb{E}\{I(T \ge u) \mid X, A = a, \delta = 1 , U \ge u\}.
\end{align}
\end{theorem}

Many common treatment effect estimands are functionals of the treatment-specific survival curves, including the survival difference at a fixed time $\tau$, the difference of RMSTs, the ratio of RMTLs, and the difference of $\tau$th quantile of survivals.
Their EIFs can be expressed in the form of a combination of weighted integrals of the EIF for treatment specific survival curves \citep{yang2020smim}. To be specific, the EIF for $\theta_{\tau}$ is 
 \begin{align}
     \varphi_{\theta_{\tau}}^{\text{eff}}(\mathcal{O}) = \int_0^{\tau}\phi_1(t)\varphi_{1}^{\text{eff}}(t;\mathcal{O})\mathrm{d}t + \int_0^{\tau}\phi_0(t)\varphi_{0}^{\text{eff}}(t;\mathcal{O})\mathrm{d}t. \label{eq:EIF ATE}
 \end{align}
 for some functions $\phi_a(\cdot)$ that $\mathbb{E}\{\phi_a(\cdot)^2\} < \infty$ (see Appendix \ref{s:asymptotic linear} for details). We limit our interest to the estimands with such form of the EIF, which covers the broad class of estimators that are favored in survival analysis.

\subsection{Augmented Calibration Weighting Estimator}
\label{sec:dacw}

Motivated by Theorem \ref{thm:EIF}, under the survival model in \eqref{eq:cox model} and the weighting models specified in \eqref{eq:sampling model}--\eqref{eq:censoring model}, we propose the following augmented CW (ACW) estimator of the treatment-specific survival curve,
\begin{align}
  \widehat{S}^{\text{ACW1}}_a(t) = &  \sum_{i=1}^N\delta_i\widehat{\omega}_i\frac{A_{ai}}{\widehat{\pi}_{ai}}e^{\widehat{\Lambda}_{ai}^C(t)}Y_{i}(t)  
 - \sum_{i=1}^N\delta_i\widehat{\omega}_i\left(\frac{A_{ai} - \widehat{\pi}_{ai}}{\widehat{\pi}_{ai}}\right)e^{-\widehat{\Lambda}_{ai}(t)} \notag\\
 &- \sum_{i=1}^N\left\{\delta_i\widehat{\omega}_i - \left(\sum_{i=1}^{N} \tildelta_i d_i\right)^{-1} \tildelta_i d_i\right\}e^{-\widehat{\Lambda}_{ai}(t)}
 + \sum_{i=1}^N\delta_i\widehat{\omega}_i\frac{A_{ai}}{\widehat{\pi}_{ai}}\int_0^t\frac{\mathrm{d}\widehat{M}_{ai}^C(u)}{e^{-\widehat{\Lambda}_{ai}^C(u)}}\frac{e^{-\widehat{\Lambda}_{ai}(t)}}{e^{-\widehat{\Lambda}_{ai}(u)}}  \notag\\
= & \sum_{i=1}^N \delta_i\widehat{\omega}_i\frac{A_{ai}}{\widehat{\pi}_{ai}} e^{\widehat{\Lambda}_{ai}^C(t)}Y_{i}(t) \notag\\
& + \sum_{i=1}^N e^{-\widehat{\Lambda}_{ai}(t)}  \left [\left(\sum_{i=1}^{N} \tildelta_i d_i\right)^{-1}\tildelta_id_i - \delta_i\widehat{\omega}_i\frac{A_{ai}}{\widehat{\pi}_{ai}} \left \{1 - \int_0^t \left \{ e^{\widehat{\Lambda}_{ai}^C(u) + \widehat{\Lambda}_{ai}(u)} \right \}\mathrm{d}\widehat{M}_{ai}^C(u) \right \} \right ].
\label{eq:ACW estimator1}
\end{align}
In addition, following the technique by \citet{zhang2012contrasting} that represents the marginal cumulative hazard function $\Lambda_a(t) = \int_0^t-\{ {S}_a(u)\}^{-1}{{\mathrm{d}S}_a(u)}$ by estimating the denominator and the numerator separately, we propose another ACW estimator,
\begin{align}
  \widehat{S}^{\text{ACW2}}_a(t) =  \exp \left \{-\int_0^t\frac{-\mathrm{d}\widehat{S}^{\text{ACW1}}_a(u)}{\widehat{S}^{\text{ACW1}}_a(u)} \right \},
  \label{eq:ACW estimator2}
 \end{align}
 where 
 \begin{align*}
     -\mathrm{d}\widehat{S}^{\text{ACW1}}_a(u) =& \sum_{i=1}^N \delta_i\widehat{\omega}_i\frac{A_{ai}}{\widehat{\pi}_{ai}} e^{\widehat{\Lambda}_{ai}^C(u)}\mathrm{d}N_{i}(u) \\
     + \sum_{i=1}^N &e^{-\widehat{\Lambda}_{ai}(u)} \mathrm{d}\widehat{\Lambda}_{ai}(u) \left [ \left(\sum_{i=1}^{N} \tildelta_i d_i\right)^{-1}\tildelta_id_i - \delta_i\widehat{\omega}_i\frac{A_{ai}}{\widehat{\pi}_{ai}} \left \{1 - \int_0^u \left \{ e^{\widehat{\Lambda}_{ai}^C(s) + \widehat{\Lambda}_{ai}(s)} \right \}\mathrm{d}\widehat{M}_{ai}^C(s) \right \} \right ]
 \end{align*}
 estimates $-\mathrm{d}S_a(u)$.
 Although $\widehat{S}^{\text{ACW1}}_a(t)$ and $\widehat{S}^{\text{ACW2}}_a(t)$ are asymptotically equivalent, simulation studies show that  $\widehat{S}^{\text{ACW2}}_a(t)$ has better finite-sample performance. 
 The proposed ACW estimators are similar to the estimators developed by \citet{zhang2012double} and \citet{zhang2019estimating}. The difference between the proposed and their method is discussed in Section \ref{sec:discussion}. 
 
 Similar to \citet{yang2020smim}, we consider an asymptotic linear characterization of the ATE estimator $\widehat{\theta}^{\text{ACW}}_{\tau} = \Psi_{\tau}\left(\widehat{S}^{\text{ACW}}_1(t), \widehat{S}^{\text{ACW}}_0(t)\right)$ for both ACW1 and ACW2 estimators. That is, under mild regularity conditions, 
 \begin{align}
     \widehat{\theta}^{\text{ACW}}_{\tau} - \theta_{\tau} = \int_0^{\tau}\phi_1(t)\left\{\widehat{S}_1^{\text{ACW}}(t) - S_1(t)\right\}\mathrm{d}t + \int_0^{\tau}\phi_0(t)\left\{\widehat{S}_0^{\text{ACW}}(t) - S_0(t)\right\}\mathrm{d}t + o_p(N^{-1/2}),
     \label{eq:linear chracterization}
 \end{align}
 for bounded variation functions $\phi_a(\cdot)$ in \eqref{eq:EIF ATE}. 
 Under the asymptotic linear characterization, $\widehat{\theta}^{\text{ACW}}_{\tau}$ has the influence function  $\varphi_{\theta_{\tau}}^{\text{eff}}(\mathcal{O})$ (see Appendix \ref{s:proof EIF ATE} for the proof).
 %that $E\{\phi_a(\cdot)^2\} < \infty$. 
 %Many survival estimands including the survival difference at a fixed time $\tau$, the difference of RMSTs, the ratio of RMTLs, and the difference of $\tau$th quantile of survivals can be represented under this asymptotic linear characterization (see Appendix \ref{s:asymptotic linear} for details). The following corollary gives the EIF of the ATE with the proof given in the Appendix \ref{s:proof EIF ATE}.
%  \begin{corollary}\label{lemma:EIF ATE}
%  Under the asymptotic linear characterization in \eqref{eq:linear chracterization}, the EIF for $\theta_{\tau}$ is 
%  \begin{align*}
%      \varphi_{\theta_{\tau}}^{\text{eff}}(\mathcal{O}) = \int_0^{\tau}\phi_1(t)\varphi_{1}^{\text{eff}}(t;\mathcal{O})\mathrm{d}t + \int_0^{\tau}\phi_0(t)\varphi_{0}^{\text{eff}}(t;\mathcal{O})\mathrm{d}t.
%  \end{align*}
%  \end{corollary}
%The corresponding ACW estimators of the ATE are $\widehat{\theta}^{ACW1}_{\tau} = \Psi_{\tau}\left(\widehat{S}^{ACW1}_1(t), \widehat{S}^{ACW1}_0(t)\right)$ and $\widehat{\theta}^{ACW2}_{\tau} = \Psi_{\tau}\left(\widehat{S}^{ACW2}_1(t), \widehat{S}^{ACW2}_0(t)\right)$.

Toward this end, the following theorem shows the local efficiency and asymptotic properties of the proposed ACW estimators of the ATE, i.e., $\widehat{\theta}^{\text{ACW1}}_{\tau} = \Psi_{\tau}\left(\widehat{S}^{\text{ACW1}}_1(t), \widehat{S}^{\text{ACW1}}_0(t)\right)$ and $\widehat{\theta}^{\text{ACW2}}_{\tau} = \Psi_{\tau}\left(\widehat{S}^{\text{ACW2}}_1(t), \widehat{S}^{\text{ACW2}}_0(t)\right)$.
\begin{theorem}\label{thm:Local efficiency}
Under Assumptions \ref{assump:treatment ignorability}--\ref{assump:OSXdis}, if either the survival model in \eqref{eq:cox model} is correctly specified or the weighting models, i.e., the sampling score model \eqref{eq:sampling model}, the treatment propensity score model \eqref{eq:propensity model}, and the censoring model \eqref{eq:censoring model}, are correctly specified, under regularity conditions, $\widehat{\theta}^{\text{ACW1}}_{\tau}$ and $\widehat{\theta}^{\text{ACW2}}_{\tau}$ are consistent for  $\theta_{\tau}$, and $N^{1/2} (\widehat{\theta}^{\text{ACW1}}_{\tau} - \theta_{\tau})$ and $N^{1/2} (\widehat{\theta}^{\text{ACW2}}_{\tau} - \theta_{\tau})$ are asymptotically normal with mean zero and variance $E(\varsigma_{1}^2)$ and $E(\varsigma_{2}^2)$.
Moreover, if all working models in \eqref{eq:cox model} and \eqref{eq:sampling model}--\eqref{eq:censoring model} are correctly specified, % under the regularity conditions, 
$\widehat{\theta}^{\text{ACW1}}_{\tau}$ and $\widehat{\theta}^{\text{ACW2}}_{\tau}$ are locally efficient, i.e., $\varsigma_{1} = \varsigma_{2} = \varphi_{\theta_{\tau}}^{\text{eff}}(\mathcal{O})$ as $n \to \infty$.
\end{theorem}
The proof of Theorem \ref{thm:Local efficiency} and details of $\varsigma_{1}^2$ and $\varsigma_{2}^2$ are provided in the Appendix \ref{s:proof asymptotic}. 
For a straightforward procedure of variance estimation, a nonparametric bootstrap method can be used. 
Specifically, we draw $B$ bootstrap samples from both the RCT and the observational sample respectively, and then for each resampled pair, we obtain a replicate of the ACW estimator; the sample variance of the $B$ bootstrap replicates is the variance of the ACW estimator.

The proposed ACW estimators depend on the parametric estimation of nuisance functions. Alternatively, we can consider a flexible nonparametric approach without the parametric assumption, which is often unrealistic in complex problems in practice. The asymptotic behavior of the ACW estimators with the nonparametric estimation of nuisance functions can be characterized by the empirical process perspective. Suppose that the posited nuisance models are consistent, i.e., $||\pi_{\delta}(X;\widehat{\eta}) - \pi_{\delta}(X)|| = o_p(1)$, $||\pi_{A}(X;\widehat{\rho}) - \pi_{A}(X)|| = o_p(1)$, and $||S_{a}(t,X;\widehat{\beta}_a) - S_a(t,X)|| = o_p(1)$,  $||S_{a}^C(t,X;\widehat{\gamma}_a) - S_a^C(t,X)|| = o_p(1)$, for $a \in \{0, 1\}$. 
Also, suppose that the weighting functions $\pi_{\delta}(X;\widehat{\eta})$, $\pi_{A}(X;\widehat{\rho})$, and $S^C_{a}(u,X;\widehat{\gamma}_a)$ are bounded away from zero.
Then, we have the effect of the estimated nuisance functions in $\widehat{\theta}^{\text{ACW}}_{\tau} - \theta_{\tau}$ bounded above by
$\sum_{a=0}^1\int_0^{\tau}\phi_a(t) \Big \{ ||\pi_{\delta}(X;\widehat{\eta}) - \pi_{\delta}(X)||\cdot||S_{a}(t,X;\widehat{\beta}_a) - S_a(t,X)|| + ||\pi_{A}(X;\widehat{\rho}) - \pi_{A}(X)||\cdot||S_{a}(t,X;\widehat{\beta}_a) - S_a(t,X)|| + \mathbb{P}  \int_0^t||\mathrm{d}{M}_a^C(u,X;\widehat{\gamma}_a)||\cdot||S_a(u,X)^{-1}S_a(t,X) - S_{a}(u,X;\widehat{\beta}_a)^{-1}S_{a}(t,X;\widehat{\beta}_a) || \Big \} \mathrm{d}t$ up to a multiplicative constant, where $\mathbb{P}$ is a true measure such that $\mathbb{P}f(\calO) = \int f(\calO) \mathrm{d}\mathbb{P}$ and $||\cdot||$ is $L_2$ norm. If each term of the bound is of rate $o_p(n^{-1/2})$ then the effect of nuisance function estimations are asymptotically negligible. This statement is formalized in the following theorem.

\begin{theorem}\label{thm:efficiency bound}
Suppose Assumptions \ref{assump:treatment ignorability}--\ref{assump:OSXdis} hold. Let $S_{a}(t,X;\widehat{\beta}_a)$ be general semiparametric and nonparametric models for $S_a(t,X)$ and $\pi_{\delta}(X;\widehat{\eta})$, $\pi_{A}(X;\widehat{\rho})$, $S_{a}^C(t,X;\widehat{\gamma}_a)$ be general semiparametric models for $\pi_{\delta}(X)$, $\pi_{A}(X)$, and $S_a^C(t,X)$, respectively, for $a \in \{0, 1\}$.
Suppose the following conditions hold: 
\begin{enumerate}
    \item[(C1)] $||S_{a}(t,X;\widehat{\beta}_a) - S_a(t,X)|| = o_p(1)$, $||\pi_{\delta}(X;\widehat{\eta}) - \pi_{\delta}(X)|| = o_p(1)$, $||\pi_{A}(X;\widehat{\rho}) - \pi_{A}(X)|| = o_p(1)$, $||S_{a}^C(t,X;\widehat{\gamma}_a) - S_a^C(t,X)|| = o_p(1)$;
    \item[(C2)] $0 < c_1 \le \pi_{\delta}(X;\widehat{\eta}), \pi_{A}(X;\widehat{\rho}), S^C_{a}(u,X;\widehat{\gamma}_a) \le c_2 \le 1$ for some $c_1, c_2$;
    \item[(C3)] $||\pi_{\delta}(X;\widehat{\eta}) - \pi_{\delta}(X)||\cdot||S_{a}(t,X;\widehat{\beta}_a) - S_a(t,X)|| = o_p(n^{-1/2})$, \\ 
    $||\pi_{A}(X;\widehat{\rho}) - \pi_{A}(X)||\cdot||S_{a}(t,X;\widehat{\beta}_a) - S_a(t,X)|| = o_p(n^{-1/2})$, \\ $\mathbb{P} \int_0^t||\mathrm{d}M_a^C(u,X;\widehat{\gamma}_a)||\cdot|| S_a(u,X)^{-1}S_a(t,X) - S_{a}(u,X;\widehat{\beta}_a)^{-1}S_{a}(t,X;\widehat{\beta}_a) || = o_p(n^{-1/2}).$ 
\end{enumerate}
Then, $\widehat{\theta}^{\text{ACW1}}_{\tau}$ and $\widehat{\theta}^{\text{ACW2}}_{\tau}$ are consistent estimators for $\theta_{\tau}$ and achieve semiparametric efficiency.
\end{theorem}
The proof of Theorem \ref{thm:efficiency bound} is provided in the Appendix \ref{s:proof efficiency bound}. To ensure the consistency of the nuisance function estimation with the convergence rate of the product as in \textit{(C3)}, we use the method of sieves in Section \ref{sec:sieve}.

\subsection{Robust Estimation using the Method of Sieves with Penalization}
\label{sec:sieve}

For a robust estimation of the ATE under possibly misspecified working models, we adopt the method of sieves \citep{grenander1981abstract, geman1982nonparametric} which enables flexible data-adaptive estimation of the survival curves and probability weights with root-n consistency \citep{chen2007large}.
We construct the sieves using the linear spans of power series \citep{newey1997convergence}, but other basis functions such as Fourier series or splines are applicable. Specifically, for a $p$-dimensional vector of non-negative integers $\kappa_k = (\kappa_{k1},...,\kappa_{kp})$, we consider a $K$-vector sieve basis functions $\bg(X) = \{ g_1(X),...,g_K(X)\}^\T = \{ X^{\kappa_1},..., X^{\kappa_K}\}^\T $, where $X^{\kappa_k} = \prod_{l = 1}^p X_l^{\kappa_{kl}}$ with $|\kappa_k| = \sum_{l=1}^p\kappa_{kl}$ non-decreasing in $k$, i.e., $|\kappa_k| \le |\kappa_{k+1}|$.
Under standard regularity conditions, the sieves approximation results in a consistent estimation of the survival curves and weighting probabilities with large $K$ (see \citealt{lee2021improving}, supporting information). To facilitate the stable estimation and to control the variability of the estimators with large $K$, we consider the penalized estimation of the nuisance functions.

For the sampling score model $\pi_{\delta}(X)$, the penalized sieves estimation is based on the dual problem of calibration that solves the estimating equation $U(\blam)$ in \eqref{eq:dualproblem}. Following the penalized estimating equation approach \citep{johnson2008penalized, wang2012penalized, yang2020doubly}, we solve
\begin{align*}
    U^{\epsilon}(\blam)=U(\blam)-q_{\epsilon}(|\blam|)\text{sign}(\blam)
\end{align*}
for $\blam=(\lambda_{1},\ldots,\lambda_{K})^{\T}$, where $q_{\epsilon}(|\blam|)\text{sign}(\blam)$ is the
element-wise product of $q_{\epsilon}(|\blam|)=\{q_{\epsilon}(|\lambda_{1}|),\ldots,q_{\epsilon}(|\lambda_{K}|)\}^{\T}$ and $\text{sign}(\blam)$. 
We define $q_{\epsilon}(x)=\mathrm{d} p_{\epsilon}(x)/\mathrm{d} x$ and specify $p_{\epsilon}(x)$ to
be the popular SCAD penalty function \citep{fan2001variable}, but different penalty functions such as adaptive lasso \citep{zou2006adaptive} or the minimax concave penalty \citep{zhang2010nearly} are also applicable. For the SCAD penalty, we have 
% \[
$q_{\epsilon}(|\lambda_{k}|)=\epsilon\left[ I(|\lambda_{k}|<\epsilon)+\{(b-1)\epsilon\}^{-1}\{(b\epsilon-|\lambda_{k}|)_{+}\}I(|\lambda_{k}|\geq\epsilon)\right]$,
% \]
for $k=1,\dots,{K}$, with $b=3.7$ following the literature and the tuning parameter $\epsilon$ selected by cross validation.

For penalized sieves estimation of the survival outcome model $S_{a}(t, X)$, the standard penalization technique for the Cox PH model was used based on the partial likelihood. Specifically, we estimate $\beta_a$ by solving 
\begin{align*}
    \arg\max_{\beta_a \in \mathbb{R}^K} \left[\sum_{r \in D} \delta_rA_{ar} \left\{\beta_a^\T \bg(X_{r})  - \log \left[\sum_{l \in R_r} \exp\{\beta_a^\T \bg(X_{l})\} \right]\right\} - \sum_{j=1}^K p_{\epsilon}(|\beta_{aj}|)\right],
\end{align*}
where $D$ is the set of indices of the events, $R_r$ is the set of indices of the patients at risk at time $t_r$, with $t_1 < ... < t_d$ denotes the distinct event time, and $p_{\epsilon}(\cdot)$ is the SCAD penalty, for $a \in \{0, 1\}$. The penalized sieves estimation of the censoring model $S_{a}^C(t, X)$ can be adopted by analogy. Lastly, for the penalized sieves estimation of the treatment propensity score model $\pi_A(X)$, we use the standard penalization approach for the binary data using the logit link with the SCAD penalty.

Under regularity conditions specified in \citet{fan2001variable} and \citet{lee2021improving}, the penalized sieve estimators of the nuisance functions possess oracle properties and satisfy two conditions in Theorem \ref{thm:efficiency bound}. Thus, the resulting ACW estimators using the method of sieves achieve the root-n consistency and semiparametric efficiency.

\section{Simulation Studies}
\label{sec:simulation}
In this section, we conduct simulation studies to evaluate the finite-sample performance of the proposed estimators. We consider the target population of size $N = 200,000$ and covariates $X =(X_1, X_2, X_3)^T$, where each $X_i, i = 1,2,3$ is generated from  $N(0,1)$ and truncated at $-4$ and $4$ to satisfy regularity conditions. 
An RCT sample of size $n \sim 1300$ is selected from a hypothetical RCT eligible population with size $N_1 = 50,000$.
From the remaining $N_2 = 150,000$ observational study population, we randomly select a sample of size $m = 5,000$.
%From a hypothetical RCT eligible population with size $N_1 = 50,000$, we generate the indicator of RCT participation according to $\log\{\pi_{\delta}(X)\} = - 4.7 - 0.5X_1 - 0.5X_2 - 0.3X_3$, which results in an RCT sample of size $n \sim 650$. From the remaining $N_2 = 150,000$ observational study population, we randomly select a sample of size $m = 5,000$.

We consider the difference in RMST as the ATE, i.e., $ \theta_{\tau} = \Psi_{\tau}(S_1(t), S_0(t)) = \int_{0}^{\tau} \left \{ S_1(t) - S_0(t) \right \}\mathrm{d}t =  \int_{0}^{\tau} S_1(t)\mathrm{d}t - \int_{0}^{\tau} S_0(t) \mathrm{d}t = \mu_{1,\tau} - \mu_{0,\tau}$, where we choose $\tau = 20$. We compare the proposed CW and the ACW estimators with four other methods, the Naive estimator based only on the RCT sample, the ZS estimator, i.e., the estimator proposed by \citet{zhang2012double} based only on the RCT sample,  %that combines the Cox PH models for event and censoring and treatment propensity score model, 
the IPSW estimator using the normalized IPSW weight in \eqref{eq:cw estimator}, and the OR estimator specified in \eqref{eq:OR estimator}. 
For the ACW estimators, in addition to the original covariate vector $g_1(X) = (X_1, X_2, X_3)^\T$, we consider the penalized sieve estimation with calibration variables $g_2(X) = (X_1,X_2,X_3,X_1X_2,X_1X_3,X_2X_3,X_1^2,X_2^2,X_3^2)^\T$, i.e., the extension of the basis functions in $g_1(X)$ to its 2nd-order power series including all two-way interactions and quadratic terms. We use (S) to indicate the method of sieves.
To assess the performance of these estimators under model misspecification, we consider four scenarios where i) all four models in \eqref{eq:cox model} and \eqref{eq:sampling model}--\eqref{eq:censoring model} are correct, ii) only the survival outcome model \eqref{eq:cox model} is correct, iii) the outcome model is incorrect but the other three weighting models \eqref{eq:sampling model}--\eqref{eq:censoring model} are correct, iv) all four models are incorrect. Details of estimators and specification of the four working models when they are correctly/incorrectly specified are listed in Table \ref{table:simulation setting}.
\arrayrulecolor{black}

\begin{table}[ht]
\caption{Simulation settings: model specification and estimators; $\mbox{expit}(x) = \{1 + \exp(-x)\}^{-1}$. O: survival outcome, S: sampling score, A: treatment propensity score, C: censoring}
\footnotesize
\label{table:simulation setting} %
    \centering  
        \resizebox{\textwidth}{!}{%
    \begin{tabular}{llll}
    \toprule
    \multicolumn{2}{l}{\textbf{Models}} & \textbf{Correctly specified} & \textbf{Incorrectly specified} \tabularnewline \cmidrule(r){1-2} \cmidrule(lr){3-3} \cmidrule(l){4-4}
    \multicolumn{1}{l}{\multirow{2}{*}{O}} & $ \lambda_1(t \mid X)$ & $t\exp(-3.7)\exp(-X_1 - X_2 - 1.5X_3)$ &  $t\exp(-0.8)\exp(- e^{X_1} - e^{X_2} - 1.5X_3)$ \tabularnewline
    & $\lambda_0(t \mid X)$ & $t\exp(-3)\exp(- 1.8X_1 - 1.5X_2 - X_3)$ & $t\exp(1.5)\exp(- 1.8e^{X_1} - 1.5e^{X_2} - X_3)$\tabularnewline
    \rowcolor{Gray}  \multicolumn{1}{c}{S} & $\pi_{\delta}(X)$ & $\expit\{- 3.9 - 0.5X_1 - 0.5X_2 - 0.3X_3\}$ & \multicolumn{1}{l}{$\mbox{expit}\{- 2.5 - 0.5e^{X_1} - 0.5e^{X_2} - 0.3X_3\}$} \tabularnewline
    \multicolumn{1}{c}{A} & $\pi_A(X)$ & 0.5 & $\mbox{expit}\{- 1 + 0.5e^{X_1} + 0.5e^{X_2} - 0.5e^{X_3}\} $ \tabularnewline
    \rowcolor{Gray}\multicolumn{1}{l}{} & $\lambda_1^C(t \mid X)$ & $t\exp(-4.5)\exp(-0.5X_1 - X_2 - X_3)$ & \multicolumn{1}{l}{$t\exp(-2.5)\exp(-0.5e^{X_1} - e^{X_2} - X_3)$}\tabularnewline 
    \rowcolor{Gray} \multicolumn{1}{l}{\multirow{-2}{*}{C}} & $\lambda_0^C(t \mid X)$ & $t\exp(-3.5)\exp(-0.5X_1 - X_2 - X_3)$ & \multicolumn{1}{l}{$t\exp(-1.5)\exp(-0.5e^{X_1} - e^{X_2} - X_3)$}\tabularnewline 
    \midrule
    \multicolumn{2}{l}{\textbf{Estimators}} & \textbf{Details} \tabularnewline \cmidrule(r){1-2} \cmidrule(l){3-4}
    \multicolumn{2}{l}{Naive} & \multicolumn{2}{l}{$\widehat{\theta}^{\text{Naive}}_{\tau} = \int_{0}^{\tau} \left \{ \widehat{S}^{\text{Naive}}_1(t) - \widehat{S}^{\text{Naive}}_0(t) \right \}\mathrm{d}t$,} \tabularnewline
    & & \multicolumn{2}{l}{where $\widehat{S}^{\text{Naive}}_a(t) =  n^{-1}\sum_{i=1}^N\delta_iA_{ai}\{\widehat{\pi}_{ai}\}^{-1}e^{\widehat{\Lambda}_{ai}^C(t)}Y_{i}(t) $} \tabularnewline
    \rowcolor{Gray} \multicolumn{2}{l}{ZS} & \multicolumn{2}{l}{The denominator of the estimator proposed by \citet{zhang2012double}} \tabularnewline
    \multicolumn{2}{l}{OR} & \multicolumn{2}{l}{$\widehat{\theta}^{\text{OR}}_{\tau} = \int_{0}^{\tau} \left \{ \widehat{S}^{\text{OR}}_1(t) - \widehat{S}^{\text{OR}}_0(t) \right \}\mathrm{d}t$, where $\widehat{S}^{\text{OR}}_a(t)$ is defined by \eqref{eq:OR estimator}} \tabularnewline
    \rowcolor{Gray} \multicolumn{2}{l}{IPSW} & \multicolumn{2}{l}{$\widehat{\theta}^{\text{IPSW}}_{\tau} = \int_{0}^{\tau} \left \{ \widehat{S}^{\text{IPSW}}_1(t) - \widehat{S}^{\text{IPSW}}_0(t) \right \}\mathrm{d}t$ where $\widehat{S}^{\text{IPSW}}_a(t) = \sum_{i=1}^N\delta_i\widehat{\omega}^{\text{IPSW}}_iA_{ai}\{\widehat{\pi}_{ai}\}^{-1}e^{\widehat{\Lambda}_{ai}^C(t)}Y_{i}(t) 
$ } \tabularnewline
    %\rowcolor{Gray} \multicolumn{2}{l}{\multirow{-2}{*}{IPSW}} & \multicolumn{2}{l}{where $\widehat{S}^{\text{IPSW}}_a(t) = \sum_{i=1}^N\delta_i\widehat{\omega}^{\text{IPSW}}_iA_{ai}\{\widehat{\pi}_{ai}\}^{-1}e^{\widehat{\Lambda}_{ai}^C(t)}Y_{i}(t) 
%$ }\tabularnewline
     \multicolumn{2}{l}{CW} & \multicolumn{2}{l}{$\widehat{\theta}^{\text{CW}}_{\tau} = \int_{0}^{\tau} \left \{ \widehat{S}^{\text{CW}}_1(t) - \widehat{S}^{\text{CW}}_0(t) \right \}\mathrm{d}t$ where $\widehat{S}^{\text{CW}}_a(t)$ is defined by \eqref{eq:cw estimator}} \tabularnewline
    \rowcolor{Gray} \multicolumn{2}{l}{ACW1} & \multicolumn{2}{l}{$\widehat{\theta}^{\text{ACW1}}_{\tau} = \int_{0}^{\tau} \left \{ \widehat{S}^{\text{ACW}}_1(t) - \widehat{S}^{\text{ACW1}}_0(t) \right \}\mathrm{d}t$ where $\widehat{S}^{\text{ACW1}}_a(t)$ is defined by \eqref{eq:ACW estimator1} with $g(X) = g_1(X)$} \tabularnewline
    \multicolumn{2}{l}{ACW2} & \multicolumn{2}{l}{$\widehat{\theta}^{\text{ACW2}}_{\tau} = \int_{0}^{\tau} \left \{ \widehat{S}^{\text{ACW2}}_1(t) - \widehat{S}^{\text{ACW2}}_0(t) \right \}\mathrm{d}t$ where $\widehat{S}^{\text{ACW2}}_a(t)$ is defined by \eqref{eq:ACW estimator2} with $g(X) = g_1(X)$} \tabularnewline
    \rowcolor{Gray}\multicolumn{2}{l}{ACW1(S)} & \multicolumn{2}{l}{The penalized ACW1 estimator using the method of sieves with $g(X) = g_2(X)$} \tabularnewline
    \multicolumn{2}{l}{ACW2(S)} & \multicolumn{2}{l}{The penalized ACW2 estimator using the method of sieves with $g(X) = g_2(X)$} \tabularnewline    
    \bottomrule
    \end{tabular}
    }
\end{table}

\begin{figure}[ht]
\centerline{
 \resizebox{\textwidth}{!}{
\includegraphics{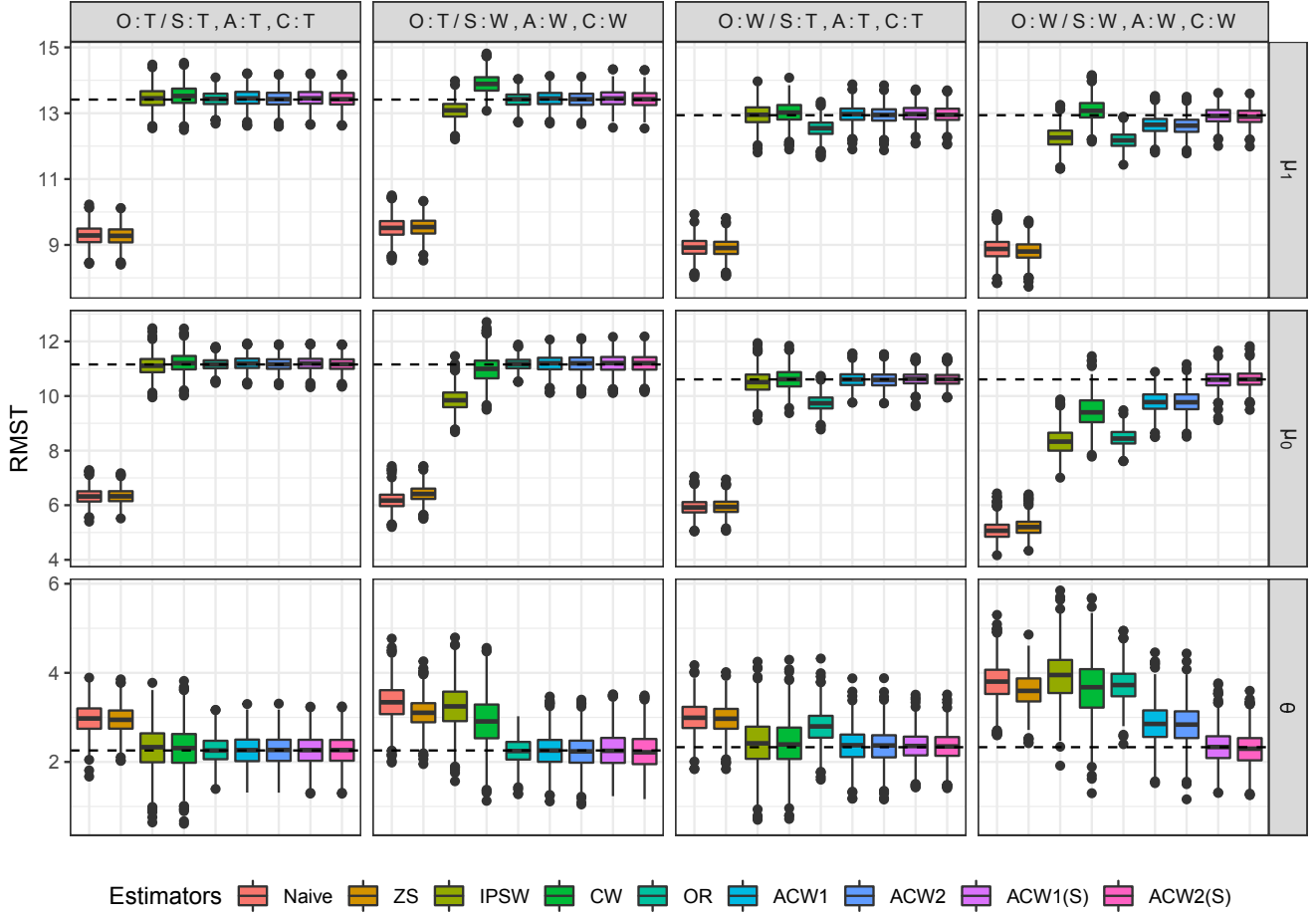}}}
\caption{Boxplot of estimators under four model specification scenarios; T: True (correct) model, W: Wrong (incorrect) model.  \label{fig:simulation}}
\end{figure}

Table \ref{table:simulation resuls} and Figure \ref{fig:simulation} summarize the results based on 1000 Monte Carlo replications. The bootstrap variance estimation was used for all estimators with $B = 100$. 
When all models are correctly specified, the Naive and ZS estimators which are based only on the RCT sample show biased estimations of the ATE due to the selection bias in the RCT sample. The OR, IPSW, CW, and ACW estimators correct that bias by leveraging the observational sample covariates. The ACW estimators were found to be more efficient than other unbiased IPW estimators. Even though the variance of the OR estimator is smaller, it is biased when the outcome model is incorrectly specified. The CW and IPSW estimators are biased when the sampling score model is not correctly specified. The proposed ACW estimators are double robust when either the outcome model or the other three models are correctly specified. In Scenario 4 where both the outcome model and the weighting models are misspecified, the ACW estimator using the penalized sieve estimation, i.e., ACW1(S) and ACW2(S), is still unbiased. The efficiency of the ACW1(S) and ACW2(S) estimators are comparable to that of the ACW1 and ACW2 estimators in Scenario 1 and 2. In Scenario 3 and 4 where the outcome model is incorrect, using the method of sieves gains efficiency. The ACW2 and ACW2(S) estimators were found to be more consistent than the ACW1 and ACW1(S) estimators in a finite sample.

\begin{table}
\centering
{\spacingset{1.25} 
\footnotesize
\caption{Simulation results under four scenarios; T: True (correct) model, W: Wrong (incorrect) model. Bias is the empirical bias of point estimates; ESE is the empirical standard error of estimates; RSE is the relative bias (\%) of bootstrap standard error estimates; CP is the empirical coverage probability of the 95\% confidence intervals.}
\label{table:simulation resuls} %
    %\resizebox{\textwidth}{!}
    {
\begin{tabular}{ccccccccccccc}
\toprule
& \multicolumn{3}{c}{\textbf{BIAS}} & \multicolumn{3}{c}{\textbf{ESE}} & \multicolumn{3}{c}{\textbf{RSE(\%)}} & \multicolumn{3}{c}{\textbf{CP(\%)}} \tabularnewline %\cmidrule(lr){3-5} \cmidrule(lr){6-8} \cmidrule(lr){9-11} \cmidrule(lr){12-14}
\cmidrule(lr){2-4} \cmidrule(lr){5-7} \cmidrule(lr){8-10} \cmidrule(lr){11-13}
 \textbf{Estimator} & $\bm{\mu_1}$ & $\bm{\mu_0}$ & $\bm{\theta}$ & $\bm{\mu_1}$ & $\bm{\mu_0}$ & $\bm{\theta}$  & $\bm{\mu_1}$ & $\bm{\mu_0}$ & $\bm{\theta}$  & $\bm{\mu_1}$ & $\bm{\mu_0}$ & $\bm{\theta}$  \tabularnewline
\midrule
%\multirow{6}{*}{\parbox{1cm}{1. O:C \\ \textcolor{white}{1.} S:C \\\textcolor{white}{1.} T:C}} 
 \multicolumn{12}{c}{{\textbf{Scenario 1: O:T / S:T, A:T, C:T}}} \vspace{1ex} \tabularnewline
Naive & -4.12 & -4.83 & 0.71 & 0.30 & 0.28 & 0.33 &  7.44 & 3.91 & 4.97 & 0.0 & 0.0 & 39.0 \tabularnewline
ZS & -4.14 & -4.83 & 0.69 & 0.29 & 0.26 & 0.30 &  7.12 & 2.13 & 4.65 & 0.0 & 0.0 & 34.2 \tabularnewline
IPSW & 0.04 & -0.04 & 0.08 & 0.32 & 0.38 & 0.49  & 2.33 & -0.14 & 2.01 & 93.9 & 94.7 & 93.3 \tabularnewline
CW & 0.11 & 0.06 & 0.05 & 0.32 & 0.37 & 0.50 & 2.57 & 1.57 & 2.29 & 92.4 & 93.9 & 93.7 \tabularnewline
OR & 0.02 & 0.01 & 0.01 & 0.22 & 0.23 & 0.30 & 1.45 & -2.59 & 1.48 & 94.7 & 95.3 & 95.0 \tabularnewline
ACW1 & 0.05 & 0.04 & 0.00 & 0.25 & 0.26 & 0.35 & 2.14 & -1.60 & 1.54 & 93.4 & 93.9 & 94.4 \tabularnewline
ACW2 & 0.02 & 0.02 & 0.00 & 0.25 & 0.26 & 0.35 & 2.27 & -0.31 & 2.56 & 94.0 & 94.5 & 94.4 \tabularnewline
ACW1(S) & 0.05 & 0.04 & 0.00 & 0.25 & 0.26 & 0.34 & 1.17 & -2.46 & -0.19 & 93.4 & 93.7 & 94.4 \tabularnewline
ACW2(S) & 0.02 & 0.02 & 0.00 & 0.25 & 0.26 & 0.35 & 1.76 & -0.98 & 1.42 & 93.9 & 94.8 & 94.1\vspace{1ex} \tabularnewline
\multicolumn{12}{c}{{\textbf{Scenario 2: O:T / S:W, A:W, C:W}}} \vspace{1ex} \tabularnewline
Naive &-3.90 & -4.98 & 1.08 & 0.31 & 0.32 & 0.39 & -0.48 & 2.62 & 1.88 & 0.0 & 0.0 & 19.5 \tabularnewline
ZS & -3.88 & -4.73 & 0.85 & 0.28 & 0.29 & 0.33 & 0.35 & 1.41 & -0.96 & 0.0 & 0.0 & 26.9 \tabularnewline
IPSW & -0.32 & -1.31 & 0.98 & 0.28 & 0.42 & 0.49 & -4.73 & 0.34 & -2.40 & 82.0 & 13.2 & 48.6 \tabularnewline
CW & 0.48 & -0.17 & 0.65 & 0.30 & 0.50 & 0.57 & -0.41 & -0.92 & -2.28 & 63.6 & 92.2 & 79.7 \tabularnewline
OR & 0.01 & 0.01 & -0.01 & 0.22 & 0.23 & 0.28 & 0.22 & -3.62 & -3.81 & 95.2 & 95.4 & 96.6 \tabularnewline
ACW1 & 0.03 & 0.03 & 0.00 & 0.24 & 0.31 & 0.38 & -0.30 & -7.07 & -5.96 & 94.3 & 95.4 & 95.4 \tabularnewline
ACW2 & 0.01 & 0.03 & -0.02 & 0.24 & 0.32 & 0.38 & -0.23 & -2.23 & -2.49 & 94.4 & 95.0 & 95.5 \tabularnewline
ACW1(S) & 0.04 & 0.03 & 0.01 & 0.26 & 0.35 & 0.42 & 0.93 & -7.55 & -7.20 & 93.0 & 94.3 & 94.6 \tabularnewline
ACW2(S) & 0.01 & 0.03 & -0.02 & 0.26 & 0.36 & 0.42 & 1.27 & -1.51 & -2.70 & 93.1 & 93.6 & 93.9\vspace{1ex} \tabularnewline
\multicolumn{12}{c}{{\textbf{Scenario 3: O:W / S:T, A:T, C:T}}} \vspace{1ex} \tabularnewline
Naive & -4.02 & -4.68 & 0.67 & 0.30 & 0.29 & 0.36 & 4.14 & 4.17 & 4.66 & 0.0 & 0.0 & 49.7 \tabularnewline
ZS & -4.03 & -4.66 & 0.63 & 0.29 & 0.28 & 0.34 & 3.23 & 3.77 & 4.88 & 0.0 & 0.0 & 49.6\tabularnewline
IPSW & 0.01 & -0.09 & 0.10 & 0.33 & 0.41 & 0.54 & 1.42 & 2.55 & 3.43 & 94.4 & 93.6 & 93.0 \tabularnewline
CW & 0.08 & 0.00 & 0.08 & 0.33 & 0.40 & 0.54 & 1.50 & 3.73 & 3.36 & 93.7 & 93.1 & 93.6 \tabularnewline
OR & -0.40 & -0.86 & 0.47 & 0.25 & 0.29 & 0.37 & -2.11 & 3.94 & 4.18 & 64.7 & 15.2 & 74.1\tabularnewline
ACW1 & 0.03 & 0.00 & 0.03 & 0.26 & 0.29  &0.37 & -0.58 & -2.68 & -1.27 & 94.3 & 93.8 & 94.7\tabularnewline
ACW2 & 0.01 & -0.01 & 0.02 & 0.26 & 0.29 & 0.37 & -0.46 & 1.54 & 1.67 & 94.5 & 93.8  & 94.5 \tabularnewline
ACW1(S) & 0.04 & 0.02 & 0.02 & 0.24 & 0.25 & 0.32 & -4.10 & -11.19 & -8.99 & 94.7 & 94.8 & 95.0 \tabularnewline
ACW2(S) & 0.02 & 0.01 & 0.01 & 0.24 & 0.25 & 0.32 & -2.19 & -1.54 & -1.50 & 95.1 & 93.9 & 94.2\vspace{1ex} \tabularnewline
\multicolumn{12}{c}{{\textbf{Scenario 4: O:W / S:W, A:W, C:W}}} \vspace{1ex} \tabularnewline
Naive & -4.06 & -5.53 & 1.48 & 0.34 & 0.33 & 0.43 & 2.18 & 5.94 & 3.48 & 0.0 & 0.0 &7.4 \tabularnewline
ZS & -4.12 & -5.40 & 1.28 & 0.31 & 0.30 & 0.38 & 2.67 & 6.78 & 4.90 & 0.0 & 0.0 & 8.0 \tabularnewline
IPSW & -0.68 & -2.27 & 1.60 & 0.32 & 0.48 & 0.55 & -2.43 & 2.19 & -2.03 & 44.1 & 1.2 & 20.2 \tabularnewline
CW & 0.14 &-1.18 & 1.33 & 0.32 & 0.56 & 0.65 & 0.59 & 1.51 & -1.80 & 92.0 & 43.0 & 46.3 \tabularnewline
OR & -0.76 & -2.15 & 1.39 & 0.25 & 0.31 & 0.38 & 2.04 & 0.54 & 2.45 & 11.7 & 0.0 & 4.9\tabularnewline
ACW1 & -0.29 & -0.82 & 0.53 & 0.26 & 0.39 & 0.45 & 2.91 & -1.69 & -2.36 & 79.4 & 43.4 & 77.1 \tabularnewline
ACW2 & -0.32 & -0.82 & 0.50 & 0.26 & 0.40 & 0.46 & 2.98 & 1.53 & 0.00 & 75.9 & 44.2 & 77.9 \tabularnewline
ACW1(S) & -0.02 & -0.02 & 0.01 & 0.24 & 0.32 & 0.38 & -0.42 & -13.16 & -11.20 & 95.1 & 96.1 & 95.4 \tabularnewline
ACW2(S) & -0.04 & 0.00 & -0.04 & 0.24 & 0.32 & 0.38 & -0.10 & -3.96 & -4.13 & 94.9 & 96.0 & 95.3
\vspace{1ex} \tabularnewline
\bottomrule 
\end{tabular}
}}

\end{table}

\newpage
\section{Real Data Application}
\label{sec:application}

We apply the proposed method to estimate the effect of adjuvant chemotherapy on survival in patients with early-stage resected non-small cell lung cancer (NSCLC). 
Cancer and Leukemia Group B (CALGB) 9633 is the only randomized phase III trial designed to evaluate the effectiveness of adjuvant chemotherapy over observation for stage IB NSCLC \citep{strauss2008adjuvant}. An additional observational sample for stage IB NSCLC patients was extracted from National Cancer Database (NCDB), including more than 15,000 patients with the same eligibility criteria as CALGB 9633. More details of the two data sources are given in \citet{lee2021improving}, where they conducted an  integrative analysis of the CALGB trial sample and the NCDB sample to improve generalizability for the CALGB 9633 trial-based estimator of average risk of cancer recurrence.

Table \ref{tab:baseline} summarizes the distribution of the four baseline covariates by the data sources, which have been considered important prognostic factors. The baseline covariates of the patients in the CALGB trial are different from those of the patients in NCDB. Specifically, the CALGB trial patients consist of more males, are younger, and have smaller tumor sizes. Consequently, an important clinical question is whether adjuvant chemotherapy benefits the general stage IB NSCLC patients population, represented by the NCDB, which is a population-based registry capturing approximately 79 percent of newly diagnosed lung cancers in the United States, and contains more females, are order, and have larger tumor sizes than the CALGB patients. Given that the RCT sample is relatively healthier than the NCDB sample, the estimators based only on CALGB 9633 sample would result in biased estimation of the true effect of adjuvant chemotherapy on the real-world population of early-stage NSCLC patients. 

\begin{table}
    \centering
    \caption{Baseline characteristics of the CALGB 9633 trial sample and the NCDB sample; mean (standard deviation) for continuous  and number (proportion) for binary covariate. }
    \label{tab:baseline}
\resizebox{\textwidth}{!}{    
    \begin{tabular}{ccccc}
    \toprule
        & Male ($X_1$)& Age ($X_2$) & Squamous histology ($X_3$) & Tumor size ($X_4$) \tabularnewline
        \midrule
        RCT: CALGB 9633 ($n = 319$) & 204 (64\%) & 60.83 (9.62) & 128 (40\%) & 4.6 (2.08)\tabularnewline
        OS: NCDB ($n = 15379$) & 8458 (55\%) & 67.87 (10.18) & 5998 (39\%) & 4.94 (3.04)\tabularnewline
        \bottomrule
    \end{tabular}
    }
\end{table}

We estimate a 12-year difference of the restricted mean lifetime between adjuvant chemotherapy and observation (i.e., no chemotherapy). The nonparametric bootstrap method is used to estimate the standard errors. The results are given in Table \ref{tab:estimated rmst}, using the proposed estimators and other existing methods in the simulation studies. The Naive and ZS estimators indicate that in the RCT sample, there is no 12-year RMST difference between the adjuvant chemotherapy and observation, i.e., $0.02$ and $0.04$ years, respectively. All other estimators that utilize the covariate information of the observational study show a much larger difference in the RMST. The IPSW and CW estimators
show about $0.4$--$0.5$ year increase in RMST for adjuvant chemotherapy over observation, and the 
OR estimator show about 0.84 year increase; however, these estimates are not significant. The proposed ACW estimators give an estimate of about a  1-year RMST increase for patients who received adjuvant chemotherapy, which is significant at 0.05 level.

\begin{figure}
\centerline{
\includegraphics[width=6in]{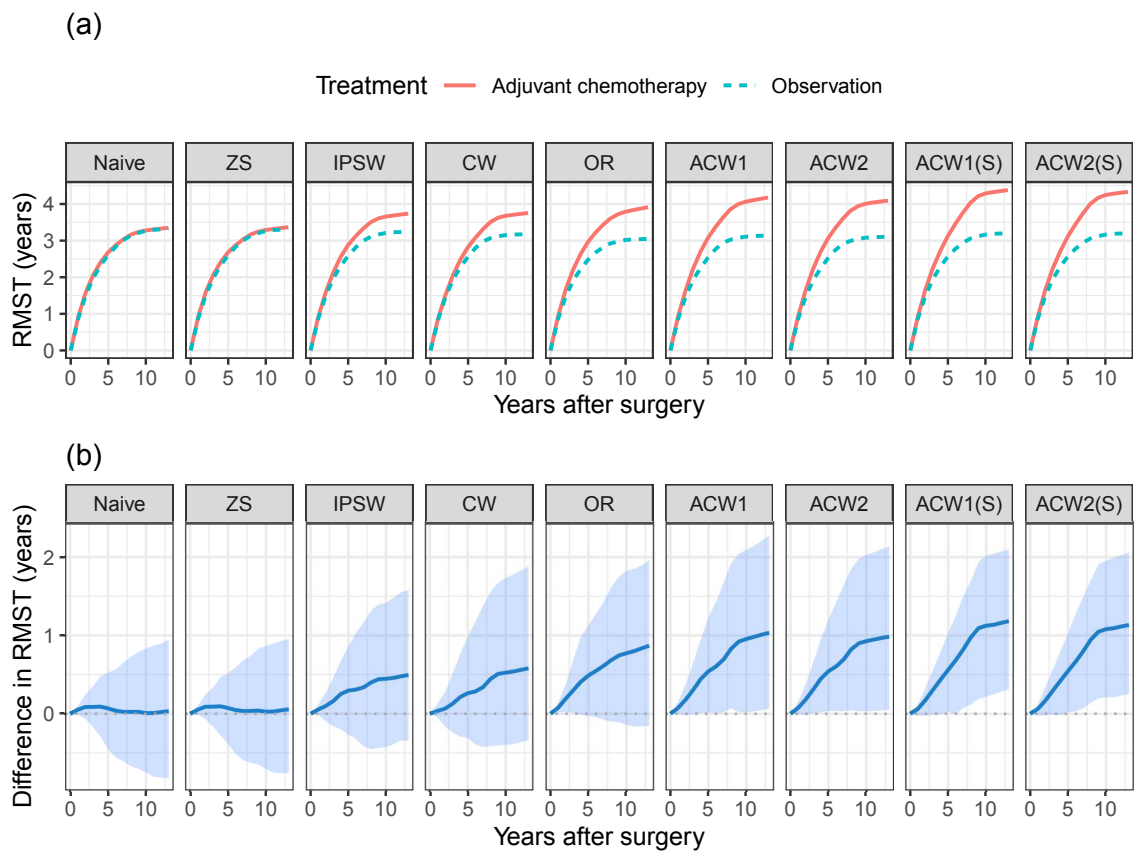}}
\caption{Estimated RMST plots of adjuvant chemotherapy and observation and their difference as a function of restricted times. \label{fig:rmst}}
\end{figure}

Figure \ref{fig:rmst} represents the estimated restricted mean lifetime for adjuvant chemotherapy and observation and their difference as a function of restricted times $\tau$. All selection-bias-adjusted estimators, i.e., the IPSW, CW, OR, ACW1, ACW2, ACW1(S), and ACW2(S) estimators, show a trend of increasing RMST difference over $\tau$ for all $\tau$ from 1 to 13 years. Especially, all of the estimators show significant non-zero differences when $\tau$ is large. Compared to the ACW1 and ACW2 estimators, the ACW1(S) and ACW2(S) estimators gain efficiency by using the method of sieves. On the other hand, the Naive and ZS estimators which are based only on the RCT sample show nearly flat trends near zero difference over $\tau$.
All these results indicate that the proposed estimators are able to detect the effect of adjuvant chemotherapy on survival in the real world population than that in the RCT sample with higher efficiency and more protection against model misspecification, by leveraging the information from the observational study.

This substantial heterogeneity in the treatment effect is mainly due to age and tumor size. While many baseline covariates are prognostic factors for survival risk, age and tumor size are the few variables associated with the outcome and significantly interact with the treatments. It is the reason that these two variables are of interest for evaluating treatment effect heterogeneity. Subgroup analysis based only on the RCT data supports the same extent of treatment effect heterogeneity varied by age and tumor size. The patients with older age and larger tumor size have significantly less risk for death after adjuvant chemotherapy than those who were younger and had smaller tumors \citep{strauss2008adjuvant}. The trend is consistent with the treatment effect heterogeneity found in the integrated analysis for the target population. The remaining difference in treatment heterogeneity could be caused by the imbalance of the covariates distribution between the RCT sample and the target population, and again the difference is expected and reasonable, and we believe it indeed reflects the value of such integrated analysis.

\begin{table}
    \centering
    \caption{Estimates and 95\% confidence intervals of 12-year difference of the restricted mean lifetime between adjuvant chemotherapy and observation. }
    \label{tab:estimated rmst}
\resizebox{0.7\textwidth}{!}{    
    \begin{tabular}{cccc}
    \toprule
       \textbf{Estimator} & $\bm{\widehat{\mu_1}}$ & $\bm{\widehat{\mu_0}}$ & $\bm{\widehat{\theta}}$  \tabularnewline
        \midrule
       Naive & 3.33 (2.86, 3.94) & 3.31 (2.81, 3.92) & 0.022 (-0.815, 0.895) \tabularnewline
       ZS & 3.34 (2.89, 3.94) & 3.30 (2.79, 3.90) &  0.044 (-0.794, 0.881) \tabularnewline
       IPSW & 3.71 (3.09, 4.55) & 3.23 (2.62, 3.83) & 0.478 (-0.339, 1.550) \tabularnewline
       CW & 3.73 (3.07, 4.69) & 3.17 (2.49, 3.87) & 0.561 (-0.363, 1.830) \tabularnewline
       OR & 3.88 (3.17, 4.69) & 3.04 (2.37, 3.61) & 0.835 (-0.160, 1.890) \tabularnewline
       ACW1 & 4.14 (3.42, 5.17) & 3.13 (2.45, 3.64) & 1.01 (0.048, 2.200)\tabularnewline
       ACW2 & 4.07 (3.35, 5.00) & 3.10 (2.41, 3.64) & 0.967 (0.043, 2.100) \tabularnewline
       ACW1(S) & 4.36 (3.56, 5.17) & 3.19 (2.66, 3.69) & 1.160 (0.290, 2.087) \tabularnewline
       ACW2(S) & 4.31 (3.50, 5.15) & 3.19 (2.66, 3.69) & 1.114 (0.239, 2.056)\tabularnewline       
        \bottomrule
    \end{tabular}
    }
\end{table}

\newpage
\section{Discussion}
\label{sec:discussion}

This paper considers a framework to estimate the treatment effect defined as a function of the treatment-specific survival probability function. The proposed ACW estimators are motivated by two identification formulas based on the survival outcome model and the weighting models and achieve local efficiency and double robustness based on semiparametric theory. 

The proposed ACW estimators are similar to the estimator proposed by \citet{zhang2012double} that combines the treatment-specific Cox model by \citet{chen2001causal} and the inverse probability of treatment weighted Nelson-Aalen method proposed by  \citet{wei2008semiparametric}. However, unlike our estimator that is derived from the EIF, their approach is based on augmenting the IPW estimating equation to achieve double robustness, and the efficiency of their estimator has not been studied yet. Moreover, their estimator is based only on the RCT sample, whereas we leverage the observational sample to account for selection bias, resulting in an additional augmentation in terms of the sampling score model. In addition, we focus on the class of estimands defined as functionals of the survival functions, including the framework proposed by \citet{zhang2012double} as a special case.
%However, unlike their estimator that is weighted only by the inverse of treatment propensity score, our estimator is additionally weighted by the inverse of sampling score model to adjust for selection bias and includes an additional augmentation in terms of the sampling score model. 
%This approach is similar to that of \citet{zhang2012double} which is based on augmenting the IPW estimating equation to achieve double robustness; however, the efficiency of their estimator has not been studied yet. We focus on the class of estimands defined as functionals of the survival functions, and show that semiparametric efficiency is achieved in our proposed framework, including the framework proposed by \citet{zhang2012double} as a special case.
The proposed approach is also similar to \citet{zhang2019estimating}, focusing on the broad class of Mann-Whitney-type causal effect based on semiparametric theory. To derive the EIF, they construct a RAL estimator that excludes censored data. In contrast, by restricting our interest to the functional of the treatment-specific survival curves, we consider a different RAL estimator utilizing both observed and censored data. The proposed class of estimators could be more robust under highly censored data and covers a broad class of estimands that are favored in time-to-event data.

Instead of a penalized nonparametric sieves method, other machine learning methods can be used, such as survival trees  \citep{bou2011review} or random forests \citep{ishwaran2008random} as alternative to the Cox proportional hazard model. 
A cross-fitting technique can be employed to remove the Donsker's condition, which is questionable to hold for these methods \citep{zhang2020efficient}.

%However, the relationship between these two approach has not been studied. 
%double-augmented semiparametric estimator for the average causal treatment effect in terms of the difference in restricted mean survival time. The proposed model combines Cox proportional hazard model, calibration weighting model, and IPTW model which allows a consistent estimation of the treatment effect in the presence of the heterogeneity between the RCT and target population, and the imbalance between the RCT treatment groups. %The proposed estimator is a consistent estimator if any two of the three working models are correctly specified. 
% We derive the asymptotic properties of the proposed estimator and evaluated its finite-sample performance via simulation.

Instead of a penalized nonparametric sieves method, other machine learning methods can be used, such as survival trees  \citep{bou2011review} or random forests \citep{ishwaran2008random} as alternative to the Cox proportional hazard model. 
A cross-fitting technique can be employed to remove the Donsker's condition, which is questionable to hold for these methods \citep{zhang2020efficient}.

%The robustness of the proposed estimators can be improved with machine learning methods. For example, a penalized Cox PH model can be used combined with a nonparametric Sieve method \citep{lee2021improving} or Bernstein polynomials \citep{osman2012nonparametric}.

 The proposed methods assume that the trial participation is ignorable, i.e., all covariates related to the trial participation and survival time are captured. However, some important covariates might not be available in the observational samples as there were not originally collected for the research purpose. Future work could involve a sensitivity analysis assessing the robustness of the proposed framework in the presence of unmeasured covariates in observational studies \citep[e.g.,][]{vanderweele2017sensitivity, yang2017sensitivity, nguyen2017sensitivity, huang2022sensitivity}.
 
In addition to the estimation of the ATE, a key challenge is to identify subgroups of patients for whom the treatment is more effective. Estimating individualized treatment effects is a key toward precision medicine so that doctors can tailor treatment for individual patients given their characteristics. Interesting future research would be generalizing individualized treatment effects for survival outcomes combining RCT and observational studies, following \citet{yang2020elastic, yang2020improved} and \citet{wu2021transfer}. 

 \bibliographystyle{Chicago}
\bibliography{ref}

%%%%%%%%%%%%%%%%%%%%%%%% Appendix %%%%%%%%%%%%%%%%%%%%%%%

\newpage{} 
\begin{center}
\textbf{\Large{}Appendix}{\Large{} }{\Large\par}
\par\end{center}

%\pagenumbering{arabic} %reset page counter to 1
%\renewcommand*{\thepage}{A\arabic{page}}

\setcounter{section}{0} 
\global\long\def\thesection{\Alph{section}}%
\setcounter{equation}{0}
\global\long\def\thesubsection{\Alph{section}.\arabic{subsection}}%
\setcounter{equation}{0} 
\global\long\def\theequation{\Alph{section}\arabic{equation}}%

\section{EIFs for survival estimands\label{s:asymptotic linear}}
All survival estimands mentioned in the main paper have the EIF in the form of a combination of weighted integrals of the EIF for treatment specific survival curves \citep{yang2020smim}. Specifically, the EIF for $\theta_{\tau}$ is 
 \begin{align*}
     \varphi_{\theta_{\tau}}^{\text{eff}}(\mathcal{O}) = \int_0^{\tau}\phi_1(t)\varphi_{1}^{\text{eff}}(t;\mathcal{O})\mathrm{d}t + \int_0^{\tau}\phi_0(t)\varphi_{0}^{\text{eff}}(t;\mathcal{O})\mathrm{d}t,
 \end{align*}
 where $\varphi_{a}^{\text{eff}}(t;\mathcal{O})$ is the EIF for treatment-specific survival curve $S_a(t)$, and $\phi_a(\cdot)$ is a function that $\mathbb{E}\{\phi_a(\cdot)^2\} < \infty$, for $a \in \{0, 1\}$.
 
 %Let $\widehat{S}_a(t)$ be a estimator of a treatment-specific survival curve $S_a(t)$ such that $\widehat{S}_a(t) - S_a(t) = o_p(N^{-1/2})$ for $a \in \{0, 1\}$.
%Following \citet{yang2020smim}, under mild regularity conditions,  following asymptotic linear characterizations,
%  \begin{align*}
%      \widehat{\theta}_{\tau} - \theta_{\tau} = \int_0^{\tau}\phi_1(t)\left\{\widehat{S}_1(t) - S_1(t)\right\}\mathrm{d}t + \int_0^{\tau}\phi_0(t)\left\{\widehat{S}_0(t) - S_0(t)\right\}\mathrm{d}t + o_p(N^{-1/2}),
%  \end{align*}
%  where $\widehat{\theta}_{\tau} = \Psi_{\tau}\left(\widehat{S}_1(t), \widehat{S}_0(t)\right)$.
 
\begin{enumerate}
    \item Difference in the survival at a time point $\tau$:\\
    $\widehat{\theta}_{\tau} = \widehat{S}_1(\tau) - \widehat{S}_0(\tau)$ $\implies \phi_1(t) = \updelta(t - \tau)$ and $\phi_0(t) = \updelta(t + \tau)$, where $\updelta(\cdot)$ is the Dirac delta function.
    \item Difference in RMSTs up to $\tau$:\\
    $\widehat{\theta}_{\tau} = \int_0^{\tau}\widehat{S}_1(\tau)\mathrm{d}t - \int_0^{\tau}\widehat{S}_0(\tau)\mathrm{d}t$ $\implies \phi_1(t) = 1$ and $\phi_0(t) = -1$.
    \item Ratio of RMTLs up to $\tau$: \\
     $\widehat{\theta}_{\tau} = \left \{ \tau - \int_0^{\tau}\widehat{S}_1(\tau)\mathrm{d}t \right \}/ \left \{\tau - \int_0^{\tau}\widehat{S}_0(\tau)\mathrm{d}t \right \}$.\\ 
     $\implies \phi_1(t) = -\left \{\tau - \int_0^{\tau}\widehat{S}_0(\tau)\mathrm{d}t \right \}^{-1}$ and $\phi_0(t) = -\widehat{\theta}_{\tau}\left \{\tau - \int_0^{\tau}\widehat{S}_0(\tau)\mathrm{d}t \right \}^{-1}$ by the Taylor expansion.
     \item Difference in $\tau$th quantile of survivals:\\
     $\theta = q_{1,\tau} - q_{0, \tau}$ and $\widehat{\theta} = \widehat{q}_{1,\tau} - \widehat{q}_{0, \tau}$ where $q_{a, \tau} = \inf_q\{S_a(q) \le \tau \}$ and $\widehat{q}_{a, \tau} = \inf_q\{\widehat{S}_a(q) \le \tau \}$.\\
     Following the Bahadur-type representation, under regularity conditions \citep{francisco1991quantile}, $\widehat{q}_{a,\tau}$ can be expressed as $$\widehat{q}_{a,\tau} - q_{a,\tau} = \frac{\widehat{S}_a(q_{a,\tau}) - S_a(q_{a,\tau})}{\dot{S}_a(q_{a,\tau})} + o_p(N^{-1/2}),$$
     where $\dot{S}_a(q_{a,\tau}) = \mathrm{d}S_a(q) / \mathrm{d}q$.\\
     $\implies \phi_1(t) = \left \{ \dot{S}_1(q_{1,\tau}) \right \}^{-1}I(t = q_{1,\tau})$ and $\phi_0(t) = \left \{ \dot{S}_0(q_{0,\tau}) \right \}^{-1}I(t = q_{0,\tau})$.
\end{enumerate}

\newpage
\section{Proofs \label{s:proofs}}

\subsection{Proof of Theorem \ref{thm:EIF} \label{s:proof EIF}}
We first derive the efficient influence function (EIF) for treatment-specific survival curves without censoring, i.e., under full data set $\mathcal{V} = (X, A, T, \delta, \tildelta)$, then derive it in the presence of censoring, i.e., under the observed data set $\mathcal{O} = (X, A, U, \Delta, \delta, \tildelta)$. The proof based on $\mathcal{V}$ is similar to the one in \citet{lee2021improving} using the method of parametric submodel \citep{bickel1993efficient}. We use $\xi$ in the subscript to denote the submodel. For example, $f_{\xi}(\V)$ is a one-dimensional parametric submodel with the true $f(\V)$ at $\xi = 0$, i.e., $f_{\xi}(\V)\rvert_{\xi=0} = f(\V)$.

Since the likelihood of a single $\V$ is \[f(\V) = \left\{f(X)f(\delta = 1 \mid X)f(A \mid X, \delta = 1)f(T \mid X, A, \delta = 1)\right\}^{\delta}\left\{f(X)\right\}^{\tildelta}\] and $\delta\tildelta = 0$, 
the score function can be decomposed as
\[\dot{\mathrm{S}}(\V) = \delta \dot{\mathrm{S}}(X) + \delta\dot{\mathrm{S}}(\delta \mid X) + \delta\dot{\mathrm{S}}(A \mid X, \delta = 1) + \delta\dot{\mathrm{S}}(T \mid X, A, \delta = 1) + \tildelta\dot{\mathrm{S}}(X).\]
We use $\dot{\mathrm{S}}$ to represent the score function.

The nuisance tangent space from the semiparametric theory is
\[H = H_1 \oplus H_2 \oplus H_3 \oplus H_4\]
where 
\begin{align*}
    H_1 &= \left \{ h(X):\mathbb{E}\{h(X)\} = 0 \right \} \\
    H_2 &= \left \{ h(X,\delta):\mathbb{E}\{h(\delta, X) \mid X\} = 0 \right \} \\
    H_3 &= \left \{ h(A, \delta, X):\mathbb{E}\{h(A, \delta, X) \mid \delta, X\} = 0 \right \} \\
    H_4 &= \left \{ h(T, A, \delta, X):\mathbb{E}\{h(T, A, \delta, X) \mid A, \delta, X\} = 0 \right \}.
\end{align*}

Here we only derive the EIF for $S_1(u)$. The EIF for $S_0(u)$ can be easily derived using the similar technique. We denote the EIF for $S_1(u)$ under $\V$ as $\varphi_1^{F}(u\mid \mathcal{V})$ which must satisfy $\varphi_1^{F}(u\mid \mathcal{V}) \in H$ and $\left.\frac{\partial}{\partial\xi} S_{1,\xi}(u) \right\rvert_{\xi = 0} = \mathbb{E}\left \{\varphi_1^{F}(u\mid \mathcal{V}) \dot{S}(\V)\right \}$ where $S_{1, \xi}(u) = \mathbb{E} \{ \tildelta d S_{1,\xi} (u, X)\}$.
Toward this end, we express
%$S_1(t \mid X) = \mathbb{E} \{I(t \ge u) \mid A=1, \delta = 1, X\}$. 
\begin{align}
    \frac{\partial}{\partial\xi} S_{1,\xi}(u) \bigg\rvert_{\xi = 0} =& \int \tildelta d S_{1,\xi} (u, X)f_{\xi}(\V)\mathrm{d}\V \bigg\rvert_{\xi = 0} \notag\\
     =& \int \tildelta d S_{1,\xi} (u, X) \dot{\mathrm{S}}_{\xi}(V)f_{\xi}(\V)\mathrm{d}\V \bigg\rvert_{\xi = 0} \notag\\
& +  \int \tildelta d \left \{\frac{\partial}{\partial\xi}S_{1,\xi} (u, X) \right \} f_{\xi}(\V)\mathrm{d}\V \bigg\rvert_{\xi = 0} \notag\\
=& \mathbb{E}\left \{  \tildelta d S_{1} (u, X) \dot{\mathrm{S}}_{\xi}(X) \right \} + \mathbb{E}\left \{ \frac{\partial}{\partial\xi}S_{1,\xi} (u, X) \bigg\rvert_{\xi = 0} \right \} \label{eq:realA1}.
\end{align} 

For the first term in \eqref{eq:realA1}, we have
\begin{align}
    & \mathbb{E}\left \{  \tildelta d S_{1} (u, X) \dot{\mathrm{S}}(X) \right \} \notag\\
    = & \mathbb{E} \left[\tildelta d \left \{ S_{1} (u, X) - S_1(u) \right \}\right]\notag\\
    = & \mathbb{E} \left[\tildelta d \left \{ S_{1} (u, X) - S_1(u) \right \} \left \{ \dot{\mathrm{S}}(X, A, T, \delta) + \tildelta\dot{\mathrm{S}}(X) \right \}\right]\notag\\
    = & \mathbb{E} \left[\tildelta d \left \{ S_{1} (u, X) - S_1(u) \right \}  \dot{\mathrm{S}}(\V) \right], \label{eq:realA2}
\end{align}
and for the second term in \eqref{eq:realA1}, we have
\begin{align}
 \frac{\partial}{\partial\xi}S_{1,\xi} (u \mid X) \bigg\rvert_{\xi = 0} &= \int I(T \ge u) \frac{\partial}{\partial\xi} f_{\xi}(u \mid X, \delta = 1, A = 1) \mathrm{d}u \bigg\rvert_{\xi = 0} \notag\\
 & = \int I(T \ge u) \dot{\mathrm{S}}_{\xi}(T \mid X, \delta = 1, A = 1)f_{\xi}(u \mid X, \delta = 1, A = 1) \mathrm{d}u \bigg\rvert_{\xi = 0} \notag\\
 & = \int I(T \ge u) \dot{\mathrm{S}}(T \mid X, \delta, A)\frac{\delta}{\pi_{\delta}(X)}\frac{A}{\pi_{A}(X)}f(t\mid X) \mathrm{d}u \notag\\
 & = \mathbb{E}\left \{ \frac{\delta}{\pi_{\delta}(X)}\frac{A}{\pi_{A}(X)} I(T \ge u) \dot{\mathrm{S}}(T \mid X, \delta, A)  \mid X \right \}.\label{eq:realA3}
\end{align}
By combining \eqref{eq:realA2} and \eqref{eq:realA3},
\begin{align*}
    \mathbb{E}\left \{ \frac{\partial}{\partial\xi}S_{1,\xi} (u, X) \bigg\rvert_{\xi = 0} \right \} &= \mathbb{E}\left \{ \frac{\delta}{\pi_{\delta}(X)}\frac{A}{\pi_{A}(X)} I(T \ge u) \dot{\mathrm{S}}(T \mid X, \delta, A)  \right \} \\
    & = \mathbb{E}\left [ \frac{\delta}{\pi_{\delta}(X)}\frac{A}{\pi_{A}(X)} \left \{ I(T \ge u) - S_1(u, X) \right \} \dot{\mathrm{S}}(T \mid X, \delta, A)  \right ] \\
    & = \mathbb{E}\left [ \frac{\delta}{\pi_{\delta}(X)}\frac{A}{\pi_{A}(X)} \left \{ I(T \ge u) - S_1(u, X) \right \} \dot{\mathrm{S}}(T, X, \delta, A)  \right ] \\
    & = \mathbb{E}\left [ \frac{\delta}{\pi_{\delta}(X)}\frac{A}{\pi_{A}(X)} \left \{ I(T \ge u) - S_1(u, X) \right \} \dot{\mathrm{S}}(\V)  \right ]. 
\end{align*}
Therefore,
\begin{align*}
    \frac{\partial}{\partial\xi} S_{1,\xi}(u) \bigg\rvert_{\xi = 0} = \mathbb{E}\left(\left [  \frac{\delta}{\pi_{\delta}(X)}\frac{A}{\pi_{A}(X)} \left \{ I(T \ge u) - S_1(u, X) \right \} + \tildelta d \left \{ S_{1} (u, X) - S_1(u) \right \} \right ] \dot{\mathrm{S}}(\V) \right).
\end{align*}

Let 
\[ \varphi_1^{F}(u\mid \mathcal{V}) = \frac{\delta}{\pi_{\delta}(X)}\frac{A}{\pi_{A}(X)} \left \{ I(T \ge u) - S_1(u, X) \right \} + \tildelta d  S_{1} (u, X) - S_1(u).\]
Since  
\[ \left \{ \tildelta d \ - \frac{\delta}{\pi_{\delta}(X)}\frac{A}{\pi_{A}(X)} \right \}S_{1} (u, X) \in H_3.\] 
and
\[  \frac{\delta}{\pi_{\delta}(X)}\frac{A}{\pi_{A}(X)} I(T \ge u) - S_1(u) \in H_4,\]
we have $\varphi_1^{F}(u\mid \mathcal{V}) \in H$, thus $\varphi_1^{F}(u\mid \mathcal{V})$ is the EIF for $S_1(u)$.

Now, in the presence of censoring, we observe data set $\mathcal{O} = (X, A, U, \Delta, \delta, \tildelta)$ instead of $\mathcal{V}$. Following \citet{robins1995analysis} and \citet{tsiatis2006semiparametric}, the EIF for the monotone coarsened data is
\begin{align*}
    & \frac{\delta}{\pi_{\delta}(X)}\frac{A}{\pi_{A}(X)}\frac{\Delta}{S_1^C(U, X)} I(U \ge u) - S_1(u) \\
    & - \frac{\delta}{\pi_{\delta}(X)}\frac{A}{\pi_{A}(X)} \left \{S_1(u, X) - \tildelta d \right \} S_{1} (u, X)  \\
    & + \int \frac{\delta}{\pi_{\delta}(X)}\frac{A}{\pi_{A}(X)} \frac{\mathrm{d}M_1^C(r, X)}{S_1^C(r, X)}g(u,r,X)
\end{align*} 
where $S_1^C(r, X) = p(C \ge r \mid X, A = 1, \delta = 1)$ and $\mathrm{d}M_1^C(r, X) = \mathrm{d}N^C(r) - \lambda_1^C(r, X)Y(r)$, $N^C(r) = I(U \le r, \Delta = 1)$ and $Y(r) = I(U \ge r)$.
According to Theorem 10.4 from \citet{tsiatis2006semiparametric}, the optimal $g(u,r,X)$ is
\[ g(u,r,X) = \mathbb{E}\left \{ I(T \ge u \mid X, A = 1, \delta = 1, T \ge r ) \right \}. \]
Thus,
\begin{align}
    & \int \frac{\delta}{\pi_{\delta}(X)}\frac{A}{\pi_{A}(X)} \frac{\mathrm{d}M_1^C(r, X)}{S_1^C(r, X)}g(u,r,X) \notag\\
    = & \int_0^{\infty} \frac{\delta}{\pi_{\delta}(X)}\frac{A}{\pi_{A}(X)} \frac{\mathrm{d}M_1^C(r, X)}{S_1^C(r, X)}\left \{ I(u > r) \frac{S_1(u, X)}{S_1(r, X)} + I(r \ge u) \cdot 1 \right \} \notag\\
    = & \int_0^u \frac{\delta}{\pi_{\delta}(X)}\frac{A}{\pi_{A}(X)} \frac{\mathrm{d}M_1^C(r, X)}{S_1^C(r, X)}\frac{S_1(u, X)}{S_1(r, X)} + \int_u^{\infty}\frac{\delta}{\pi_{\delta}(X)}\frac{A}{\pi_{A}(X)}\frac{\mathrm{d}M_1^C(r, X)}{S_1^C(r, X)}.
    \label{eq:realA4}
\end{align}
For the second term in \eqref{eq:realA4}, we can express as
\begin{align}
    \eqref{eq:realA4} = & \frac{\delta}{\pi_{\delta}(X)}\frac{A}{\pi_{A}(X)}\left \{ \int_u^{\infty} \frac{ \mathrm{d}N^C(r)}{S_1^C(r, X)} -  \int_u^{\infty} \frac{\lambda_1(r, X)Y(r)}{S_1^C(r, X)} \right \} \notag\\
    = & \frac{\delta}{\pi_{\delta}(X)}\frac{A}{\pi_{A}(X)}Y(u) \left \{ \frac{1 - \Delta}{S_1^C(U, X)} - \int_u^U \frac{\lambda_1(r, X)}{S_1^C(r, X)}\right \} \notag\\
   = & \frac{\delta}{\pi_{\delta}(X)}\frac{A}{\pi_{A}(X)}Y(u) \left [\frac{1 - \Delta}{S_1^C(U, X)} - \left \{ \frac{1}{S_1^C(U, X)} - \frac{1}{S_1^C(u, X)} \right \} \right ] \notag\\
   = & \frac{\delta}{\pi_{\delta}(X)}\frac{A}{\pi_{A}(X)}Y(u) \left \{ \frac{1}{S_1^C(u, X)} - \frac{\Delta}{S_1^C(U, X)} \right \}. \label{eq:realA5}
\end{align}
 Plugging \eqref{eq:realA5} to \eqref{eq:realA4}, the EIF for $S_1(u)$ under $\calO$ is
 \begin{align*}
    & \frac{\delta}{\pi_{\delta}(X)}\frac{A}{\pi_{A}(X)}\frac{\Delta}{S_1^C(U, X)} I(U \ge u) - S_1(u) \\
    & - \frac{\delta}{\pi_{\delta}(X)}\frac{A}{\pi_{A}(X)} \left \{S_1(u, X) - \tildelta d \right \} S_{1} (u, X)  \\
    & + \int_0^u \frac{\delta}{\pi_{\delta}(X)}\frac{A}{\pi_{A}(X)} \frac{\mathrm{d}M_1^C(r, X)}{S_1^C(r, X)}\frac{S_1(u, X)}{S_1(r, X)} \\
    & - \frac{\delta}{\pi_{\delta}(X)}\frac{A}{\pi_{A}(X)}Y(u) \left \{ \frac{1}{S_1^C(u, X)} - \frac{\Delta}{S_1^C(U, X)} \right \} \\
    = & \frac{\delta}{\pi_{\delta}(X)}\frac{A}{\pi_{A}(X)}\frac{I(U \ge u)}{S_1^C(u, X)} - S_1(u) \\
    & - \frac{\delta}{\pi_{\delta}(X)}\frac{A}{\pi_{A}(X)} \left \{S_1(u, X) - \tildelta d \right \} S_{1} (u, X)  \\
    & + \int_0^u \frac{\delta}{\pi_{\delta}(X)}\frac{A}{\pi_{A}(X)} \frac{\mathrm{d}M_1^C(r, X)}{S_1^C(r, X)}\frac{S_1(u, X)}{S_1(r, X)}.
 \end{align*}
 The EIF for $S_0(t)$ can be obtained by analogy.

\subsection{The influence function of the ACW estimators \label{s:proof EIF ATE}}
%\sy{subsection title?}
For both ACW1 and ACW2 estimators, by the definition of the EIF, 
$$\sqrt{N}\left \{\widehat{S}^{\text{ACW}}_a(t) - S_a(t)\right \} = \frac1{\sqrt{N}}\sum_{i=1}^N\varphi_{a}^{\text{eff}}(t;\calO_i) + o_p(1), \mbox{ for } a \in \{0, 1\}.$$
Thus, under asymptotic linear characterization of the ATE estimator $\widehat{\theta}^{\text{ACW}}_{\tau} = \Psi_{\tau}\left(\widehat{S}^{\text{ACW}}_1(t), \widehat{S}^{\text{ACW}}_0(t)\right)$, 
\begin{align*}
  \sqrt{N} (\widehat{\theta}^{\text{ACW}}_{\tau} - \theta_{\tau}) &= \int_0^{\tau}\phi_1(t)\sqrt{N}\left\{\widehat{S}_1^{\text{ACW}}(t) - S_1(t)\right\}\mathrm{d}t + \int_0^{\tau}\phi_0(t)\sqrt{N}\left\{\widehat{S}_0^{\text{ACW}}(t) - S_0(t)\right\}\mathrm{d}t + o_p({1}) \\
  & = \int_0^{\tau}\phi_1(t)\frac1{\sqrt{N}}\sum_{i=1}^N\varphi_{1}^{\text{eff}}(t;\calO_i)\mathrm{d}t + \int_0^{\tau}\phi_0(t)\frac1{\sqrt{N}}\sum_{i=1}^N\varphi_{0}^{\text{eff}}(t;\calO_i)\mathrm{d}t + o_p({1})\\
  &= \frac1{\sqrt{N}} \sum_{i=1}^N \left \{\int_0^{\tau} \phi_1(t)\varphi_{1}^{\text{eff}}(t;\calO_i)\mathrm{d}t + \int_0^{\tau}\phi_0(t)\varphi_{0}^{\text{eff}}(t;\calO_i) \mathrm{d}t \right \}+ o_p({1}).
\end{align*}
Thus, $\widehat{\theta}^{\text{ACW}}_{\tau}$ has the influence function
 \begin{align*}
     \varphi_{\theta_{\tau}}^{\text{eff}}(\calO) = \int_0^{\tau}\phi_1(t)\varphi_{1}^{\text{eff}}(t;\calO)\mathrm{d}t + \int_0^{\tau}\phi_0(t)\varphi_{0}^{\text{eff}}(t;\calO)\mathrm{d}t.
 \end{align*}

\subsection{Proof of Theorem \ref{thm:Local efficiency} \label{s:proof asymptotic}}

Standard regularity conditions for the consistency of survival outcome regression parameters and treatment propensity score model \citep[e.g.,][]{lin1989robust,zhang2012double}, as well as suitable regularity conditions for the M-estimation theory \citep[e.g.,][]{van2000asymptotic} are assumed throughout this proof. The former includes almost sure boundedness of $X$ and finite cumulative baseline hazard function for survival and censoring, and the latter includes smoothness and identifiability of working models and applicability of dominated convergence theorem.

Under Assumptions \ref{assump:treatment ignorability}--\ref{assump:noninformative censoring}, we consider three cases; (i) survival outcome regression model (O) in \eqref{eq:cox model} is correctly specified, (ii) the weighting models, i.e., the sampling score model (S) in \eqref{eq:sampling model}, the treatment propensity score model (A) in \eqref{eq:propensity model}, and the censoring model (C) in \eqref{eq:censoring model}, are correctly specified, (iii) all working models in \eqref{eq:cox model} and \eqref{eq:sampling model}--\eqref{eq:censoring model} are correctly specified. Assume that $\widehat{\zeta} = (\widehat{\eta}, \widehat{\rho}, \widehat{\beta}_1, \widehat{\beta}_0, \widehat{\gamma}_1, \widehat{\gamma}_0)$ converges to some $\zeta^* = (\eta^*, \rho^*, \beta^*_1, \beta^*_0,  \gamma^*_1, \gamma^*_0)$, not necessary true. 
Then, $\widehat{\pi}_{\delta}(X)=\pi_{\delta}(X ; \widehat{\eta}) = \exp\{\widehat{\eta}^{T}\bg(X)\}$ converges to $\pi_{\delta}^*(X)$, $\widehat{\pi}_{A}(X) = \pi_{A}(X; \widehat{\rho})=\left[1 + \exp\{-\widehat{\rho}^{T}\bg(X)\} \right]^{-1}$ converges to $\pi^*_{A}(X)$,
$\widehat{\Lambda}_a(t) = \widehat{\Lambda}_a(t;\widehat{\beta}_a)$ converges to $\Lambda^*_a(t)$, and $\widehat{\Lambda}_{a}^C(t) = \widehat{\Lambda}_{a}^C(t; \widehat{\gamma}_a)$ converges to $\Lambda_{a}^{*C}(t)$, for $a \in \{0, 1\}$.
The following proof is similar to the one in \citet{zhang2012double} and \citet{zhang2019estimating}.

\subsubsection*{Double Robustness}
We first demonstrate the double robustness property of $\widehat{S}^{\text{ACW1}}_1(t)$. 
It is straightforward to show double robustness of $\widehat{S}^{\text{ACW1}}_0(t)$ and $\widehat{\theta}^{\text{ACW1}}$ by analogy.
For the simplicity, we only consider the simple random sampling in the observational study, but it can be easily extended to the general setting with known design weights.
Define $\mathcal{S}_{1}(t;\zeta^*) = \{\pi^*_{\delta}(X) \pi^*_{A}(X) e^{-\Lambda_{a}^{*C}(t)} \}^{-1} \delta AY(t)
            - \{\pi^*_{\delta}(X) \pi^*_A(X) \}^{-1} \delta \{A - \pi^*_A(X) \}  e^{-\Lambda_{1}^{*}(t)}
            - \left\{ \pi^*_{\delta}(X)^{-1} \delta - \tildelta d \right\} e^{-\Lambda_{1}^{*}(t)} 
            + \{\pi^*_{\delta}(X)\pi^*_{A}(X)\}^{-1} \delta A \int_0^t \{e^{-\Lambda_{1}^{*C}(t)}e^{-\Lambda_{1}^{*}(u)}\}^{-1} e^{-\Lambda_{1}^{*}(t)}\mathrm{d}M^{*C}_1(u)$.
Under mild regularity conditions, $\widehat{S}^{\text{ACW1}}_1(t) \overset{p}{\longrightarrow} \mathbb{E}\left\{\mathcal{S}_{1}(t;\zeta^*)\right\}$ where 
\begin{align}
    \mathbb{E}\left\{\mathcal{S}_{1}(t;\zeta^*)\right\} = & \mathbb{E}\left\{\frac{\delta}{\pi^*_{\delta}(X)}\frac{A}{\pi^*_{A}(X)}\frac{Y(t)}{e^{-\Lambda_{a}^{*C}(t)}}\right\} \label{eq:A1} \\
            & - \mathbb{E}\left[ \frac{\delta}{\pi^*_{\delta}(X)}\left\{\frac{A - \pi^*_A(X)}{\pi^*_A(X)}\right\}  e^{-\Lambda_{1}^{*}(t)}\right] \label{eq:A2}\\
            & - \mathbb{E}\left[ \left\{ \frac{\delta}{\pi^*_{\delta}(X)} - \tildelta d \right\} e^{-\Lambda_{1}^{*}(t)} \right] \label{eq:A3}\\
            & + \mathbb{E}\left\{ \frac{\delta}{\pi^*_{\delta}(X)}\frac{A}{\pi^*_{A}(X)}\int_0^t \frac{\mathrm{d}M^{*C}_1(u)}{e^{-\Lambda_{1}^{*C}(t)}}\frac{e^{-\Lambda_{1}^{*}(t)}}{e^{-\Lambda_{1}^{*}(u)}}\right\} \label{eq:A4}.
\end{align}

First, consider the condition (i) that the model for O is correct.
Using the notation $\kappa(t)$ from \citet{zhang2012double}, we can express $Y(t) = I(T \ge t)\kappa(t)$ where $\kappa(t) = I(C \ge T \mbox{ or } C \ge t)$. 
Following \citet[Chapter 9.3 and Lemma 10.4]{tsiatis2006semiparametric}, we can write
\[ \frac{\kappa(t)}{e^{-\Lambda_{1}^{*C}(t)}} = 1 - \int_0^t \frac{\mathrm{d}M_1^{*C}(u)}{e^{-\Lambda_{1}^{*C}(t)}}.\]
Then we have
\begin{align}
    \eqref{eq:A1}  = & \mathbb{E}\left\{\frac{\delta}{\pi^*_{\delta}(X)}\frac{A}{\pi^*_{A}(X)}\frac{I(T \ge t)\kappa(t)}{e^{-\Lambda_{1}^{*C}(t)}}\right\} \notag\\
      = & \mathbb{E}\left\{\frac{\delta}{\pi^*_{\delta}(X)}\frac{A}{\pi^*_{A}(X)} I(T \ge t) - \frac{\delta}{\pi^*_{\delta}(X)}\frac{A}{\pi^*_{A}(X)}\int_0^t \frac{\mathrm{d}M_1^{*C}(u)}{e^{-\Lambda_{1}^{*C}(t)}}I(T \ge t) \right\} \label{eq:A5} \\
     \eqref{eq:A2} + \eqref{eq:A3} = & 
             - \mathbb{E}\left\{ \frac{\delta}{\pi^*_{\delta}(X)} \frac{A}{\pi^*_A(X)} e^{-\Lambda_{1}^{*}(t)}\right\} \label{eq:A6} \\
             & + \mathbb{E}\left\{ \tildelta d e^{-\Lambda_{1}^{*}(t)} \right\}. \label{eq:A7}
\end{align}
Combining \eqref{eq:A4} with \eqref{eq:A5}--\eqref{eq:A7}, we have
\begin{align}
    & \mathbb{E}\left\{ \tildelta d e^{-\Lambda_{1}^{*}(t)} \right\}  \label{eq:A8} \\
    & + \mathbb{E}\left[\frac{\delta}{\pi^*_{\delta}(X)}\frac{A}{\pi^*_{A}(X)} \left \{I(T \ge t) - e^{-\Lambda_{1}^{*}(t)} \right\} \right] \label{eq:A9}\\
    & + \mathbb{E}\left[ \frac{\delta}{\pi^*_{\delta}(X)}\frac{A}{\pi^*_{A}(X)}\int_0^t \frac{\mathrm{d}N^{C}_1(u)}{e^{-\Lambda_{1}^{*C}(t)}}\left\{ I(T \ge t) - \frac{e^{-\Lambda_{1}^{*}(t)}}{e^{-\Lambda_{1}^{*}(u)}}\right\} \right] \label{eq:A10}\\
    & - \mathbb{E}\left[ \frac{\delta}{\pi^*_{\delta}(X)}\frac{A}{\pi^*_{A}(X)}\int_0^t \frac{Y(u)\mathrm{d}\Lambda^{*C}_1(u)}{e^{-\Lambda_{1}^{*C}(t)}}\left\{ I(T \ge t) - \frac{e^{-\Lambda_{1}^{*}(t)}}{e^{-\Lambda_{1}^{*}(u)}}\right\} \right] \label{eq:A11}.    
\end{align}

Since $e^{-\Lambda_{1}^{*}(t)} = p(T \ge t \mid X, A = 1, \delta = 1)$, \eqref{eq:A8} equals to $S_1(t)$ and  \eqref{eq:A9} is  $0$ by iterated expectation conditioning on $(X, A = 1, \delta = 1)$. Also \eqref{eq:A10} and \eqref{eq:A11} are 0 by iterated expectations conditioning on $(X, A = 1, \delta = 1, C = u, T \ge u)$ and $(X, A = 1, \delta = 1, C \ge u, T \ge u)$, respectively.

Now, consider the condition (ii) that the models for S, A, and C are correct. 
As $\pi^*_{\delta}(X) = p(\delta = 1 \mid X)$, $\pi^*_{A}(X) = p(A = 1 \mid X, \delta = 1)$, and  $e^{-\Lambda_{1}^{*C}(t)} = p(C \ge t \mid X, A = 1, \delta = 1)$,
\begin{align*}
    \eqref{eq:A1} = & \mathbb{E}\left[\frac{\delta}{\pi^*_{\delta}(X)}\mathbb{E} \left \{\frac{A}{\pi^*_{A}(X)}\frac{I(T \ge t)I(C\ge t)}{e^{-\Lambda_{a}^{*C}(t)}} \bigg \rvert X, \delta = 1 \right \}\right] \\
     = & \mathbb{E}\left\{\frac{\delta}{\pi^*_{\delta}(X)}\frac{\pi^*_{A}(X)}{\pi^*_{A}(X)}\frac{P(T \ge t \mid X,A = 1, \delta = 1)P(C\ge t\mid X,A = 1, \delta = 1)}{e^{-\Lambda_{a}^{*C}(t)}} \right\} \\
     = & \mathbb{E}\left\{\tildelta d P(T \ge t \mid X,A = 1, \delta = 1)\right\} \\
     = & S_1(t)\\
     \eqref{eq:A2} + \eqref{eq:A3} = & 
             \mathbb{E}\left[ e^{-\Lambda_{1}^{*}(t)} \left\{ \tildelta d - \frac{\delta}{\pi^*_{\delta}(X)} \frac{\pi^*_{A}(X)}{\pi^*_A(X)} \right\} \right ] = 0 \\
    \eqref{eq:A4} = &  \mathbb{E}\left[ \frac{\delta}{\pi^*_{\delta}(X)}\frac{\pi^*_{A}(X)}{\pi^*_{A}(X)} \mathbb{E} \left\{\int_0^t \frac{\mathrm{d}M^{*C}_1(u)}{e^{-\Lambda_{1}^{*C}(t)}}\frac{e^{-\Lambda_{1}^{*}(t)}}{e^{-\Lambda_{1}^{*}(u)}} \rvert X, A=1, \delta = 1\right\} \right] = 0.
\end{align*}

Thus, $\widehat{S}^{\text{ACW1}}_0(t)$ has the double robustness property. By analogy, it is straightforward to show that $\widehat{S}^{\text{ACW1}}_0(t)$ converges to $\mathbb{E}\left\{\mathcal{S}_{0}(t; \zeta^*)\right\}$ which equals to $S_0(t)$ if either working model for O or weighting models S, A, and C are correct. Consequently, $\widehat{\theta}^{\text{ACW1}}_\tau$ converges to $\mathbb{E}\left\{\vartheta_{\tau}(\zeta^*)\right\}$, where
\[\vartheta_{\tau}(\zeta^*) = \int_0^{\tau}\phi_1(t)\mathcal{S}_{1}(t;\zeta^*)\mathrm{d}t + \int_0^{\tau}\phi_0(t)\mathcal{S}_{0}(t;\zeta^*)\mathrm{d}t.\]
Then, $\mathbb{E}\left\{\vartheta_{\tau}(\zeta^*)\right\}$ equals to $\theta_{\tau}$ with the same double robustness properties.

For the ACW2 estimator, we can show that the numerator $-\mathrm{d}\widehat{S}^{\text{ACW1}}_1(u)$ converges in probability to $-\mathrm{d}S_1(u)$, thus $\widehat{S}_1^{\text{ACW2}}(t) \overset{p}{\longrightarrow} S_1(t)$. Similarly, one can easily obtain $\widehat{S}_0^{\text{ACW2}}(t) \overset{p}{\longrightarrow} S_0(t)$ and $\widehat{\theta}^{\text{ACW2}}_\tau \overset{p}{\longrightarrow} \theta_\tau$.

\subsubsection*{Asymptotic Normality}
Following empirical process literature, define $\mathbb{P}$ as the true measure, $\mathbb{P}_N$ as the empirical measure, and define $\mathbb{G}_N = \sqrt{N}(\mathbb{P}_N - \mathbb{P})$ for the empirical processes.
Following the technique used in \citet{zhang2019estimating}, we assume that $\zeta$ and  $\zeta^*$ takes values in a suitable Banach space with
\begin{align}
    \sqrt{N}(\widehat{\eta} - \eta^*) & = \mathbb{G}_N\boldsymbol{\varphi}_{\eta}(\mathcal{O}) + o_p(1)\notag\\
    \sqrt{N}(\widehat{\rho} - \rho^*) & = \mathbb{G}_N\boldsymbol{\varphi}_{\rho}(\mathcal{O}) + o_p(1)\notag\\
    \sqrt{N}(\widehat{\beta}_a - \beta_a^*) & = \mathbb{G}_N\boldsymbol{\varphi}_{\beta_a}(\mathcal{O}) + o_p(1)\notag\\
    \sqrt{N}(\widehat{\gamma}_a - \gamma_a^*) & = \mathbb{G}_N\boldsymbol{\varphi}_{\gamma_a}(\mathcal{O}) + o_p(1),
\end{align}
for $a \in \{0, 1\}$.
%  According to \citet{lin1989robust}, under regularity conditions, $\sqrt{N}(\widehat{\beta} - \beta^*)$ is asymptotically normal with
% \begin{align}
%     \sqrt{N}(\widehat{\beta}_a - \beta_a^*) = \Omega_a^{-1}(\beta_a^*)N^{1/2}\sum_{i=1}^NU_{\beta_a,i}(\beta_a^*) + o_p(1)
% \end{align}
% where $U_{\beta_a,i}(\beta_a^*) = $
Assuming that $\vartheta_{\tau}(\zeta^*)$ belongs to Donsker classes \citep{van1996weak, kennedy2016semiparametric}, and under assumed regularity conditions, $\mathbb{P}_N\vartheta_{\tau}(\widehat{\zeta}) = \widehat{\theta}_{\tau}^{\text{ACW1}}$ and $\mathbb{P}\vartheta_{\tau}(\zeta^*) = \theta_{\tau}$.
Then,
\begin{align}
     \sqrt{N}\left\{\widehat{\theta}_{\tau}^{\text{ACW1}} - \theta_{\tau} \right\} = & \sqrt{N} \left \{\mathbb{P}_N\vartheta_{\tau}(\widehat{\zeta}) -  \mathbb{P}\vartheta_{\tau}(\zeta^*)\right \} \notag\\
     = & \mathbb{G}_N\vartheta_{\tau}(\widehat{\zeta}) + \sqrt{N} \mathbb{P}\left \{\vartheta_{\tau}(\widehat{\zeta}) - \vartheta_{\tau}(\zeta^*)\right \}\notag\\
     = & \mathbb{G}_N\vartheta_{\tau}(\zeta^*) + \sqrt{N} \mathbb{P}\left\{\vartheta_{\tau}(\widehat{\zeta}) - \vartheta_{\tau}(\zeta^*)\right \} + o_p(1). \label{eq:effbound}
\end{align}
As $\vartheta$ is differentiable at $\zeta^*$, we have the derivative
\begin{align*}
\mathrm{D}^{\text{ACW}} = (\mathrm{D}_{\eta}^{\text{ACW}}, \mathrm{D}_{\rho}^{\text{ACW}}, \mathrm{D}_{\beta_1}^{\text{ACW}}, \mathrm{D}_{\beta_0}^{\text{ACW}},  \mathrm{D}_{\gamma_1}^{\text{ACW}}, \mathrm{D}_{\gamma_0}^{\text{ACW}}) 
= \int_0^{\tau}\left \{ \phi_1(t)\mathrm{D}\mathcal{S}_1 + \phi_0(t)\mathrm{D}\mathcal{S}_0 \right \}\mathrm{d}t.    
\end{align*}
This implies that
\begin{align*}
    & \sqrt{N} \mathbb{P}\left\{\vartheta_{\tau}(\widehat{\zeta}) - \vartheta_{\tau}(\zeta^*)\right \}  \\
    = & \mathrm{D}^{\text{ACW}}\sqrt{N}(\widehat{\eta} - \eta^*, \widehat{\rho} - \rho^*, \widehat{\beta}_1 - \beta_1^*, \widehat{\beta}_0 - \beta_0^*,  \widehat{\gamma}_1 - \gamma_1^*, \widehat{\gamma}_0 - \gamma_0^*) + o_p(1) \\
    = & \mathbb{G}_N(\mathrm{D}_{\eta}^{\text{ACW}}\boldsymbol{\varphi}_{\eta} + \mathrm{D}_{\rho}^{\text{ACW}} \boldsymbol{\varphi}_{\rho} +  \mathrm{D}_{\beta_1}^{\text{ACW}}\boldsymbol{\varphi}_{\beta_1} + \mathrm{D}_{\beta_0}^{\text{ACW}}\boldsymbol{\varphi}_{\beta_0} + \mathrm{D}_{\gamma_1}^{\text{ACW}}\boldsymbol{\varphi}_{\gamma_1}+\mathrm{D}_{\gamma_0}^{\text{ACW}}\boldsymbol{\varphi}_{\gamma_0}) + o_p(1).
\end{align*}
Therefore, 
\begin{align*}
    \sqrt{N} \left \{\widehat{\theta}_{\tau}^{\text{ACW1}} - \theta_{\tau} \right \} = & \mathbb{G}_N\left \{\vartheta_{\tau}(\zeta^*) + \mathrm{D}_{\eta}^{\text{ACW}}\boldsymbol{\varphi}_{\eta} + \mathrm{D}_{\rho}^{\text{ACW}} \boldsymbol{\varphi}_{\rho} +  \mathrm{D}_{\beta_1}^{\text{ACW}}\boldsymbol{\varphi}_{\beta_1} + \mathrm{D}_{\beta_0}^{\text{ACW}}\boldsymbol{\varphi}_{\beta_0} + \mathrm{D}_{\gamma_1}^{\text{ACW}}\boldsymbol{\varphi}_{\gamma_1} + \mathrm{D}_{\gamma_0}^{\text{ACW}}\boldsymbol{\varphi}_{\gamma_0} \right \} \\
    & + o_p(1).
\end{align*}
The similar technique can be used to prove the asymptotic normality of the ACW2 estimator using the Delta method and the M-estimation theory.

\subsubsection*{Local Efficiency}
Define the asymptotic variance of the ACW1 estimator as $\mathbb{E}(\varsigma_1^2)$, where
\[\varsigma_1 = \vartheta_{\tau}(\zeta^*) + \mathrm{D}_{\eta}^{\text{ACW}}\boldsymbol{\varphi}_{\eta} + \mathrm{D}_{\rho}^{\text{ACW}} \boldsymbol{\varphi}_{\rho} +  \mathrm{D}_{\beta_1}^{\text{ACW}}\boldsymbol{\varphi}_{\beta_1} + \mathrm{D}_{\beta_0}^{\text{ACW}}\boldsymbol{\varphi}_{\beta_0} + \mathrm{D}_{\gamma_1}^{\text{ACW}}\boldsymbol{\varphi}_{\gamma_1} + \mathrm{D}_{\gamma_0}^{\text{ACW}}\boldsymbol{\varphi}_{\gamma_0}.\]
When the model for O is correct, then $\mathrm{D}_{\beta_1}^{\text{ACW}} = 0$ and $\mathrm{D}_{\beta_0}^{\text{ACW}} = 0$. When the weighting models S, A, and C are correct then 
$\mathrm{D}_{\eta}^{\text{ACW}}, \mathrm{D}_{\rho}^{\text{ACW}}, \mathrm{D}_{\gamma_1}^{\text{ACW}}, \mathrm{D}_{\gamma_0}^{\text{ACW}}$ are all zero. Thus, when all four working models are correct, then 
\[\sqrt{N}\left \{ \widehat{\theta}_{\tau}^{\text{ACW1}} - \theta_{\tau} \right \}= \mathbb{G}_N\vartheta_{\tau}(\zeta^*) + o_p(1) = \mathbb{G}_N\varphi_{\theta_{\tau}}^{\text{eff}}(\calO) + o_p(1),
\]
i.e., $\varsigma_1 = \varphi_{\theta_{\tau}}^{\text{eff}}(\calO)$, implying that the ACW1 estimator is the locally efficient estimator. The efficiency of the ACW2 estimator can be proved using the similar technique and the small order difference between the ACW1 and the ACW2 estimator.

\subsection{Proof of Theorem \ref{thm:efficiency bound} \label{s:proof efficiency bound}}
From \eqref{eq:effbound} in \ref{s:proof asymptotic}, we want to show that $\mathbb{P}\left\{\vartheta_{\tau}(\widehat{\zeta}) - \vartheta_{\tau}(\zeta^*)\right \}$ is negligible under conditions in Theorem \ref{thm:efficiency bound}. Toward this end, we write 
\begin{align*}
    \mathbb{P}\left\{\vartheta_{\tau}(\widehat{\zeta}) - \vartheta_{\tau}(\zeta^*)\right \} &= \sum_{a=0}^1 \int_0^{\tau}\phi_a(t)\mathbb{P}\left\{\mathcal{S}_{a}(t; \widehat{\zeta}) - \mathcal{S}_{a}(t; \zeta^*)\right\}\mathrm{d}t,
\end{align*}
where $\mathcal{S}_{a}(t; \widehat{\zeta})$ is defined in \ref{s:proof asymptotic}.
As $Y(t) = I(T \ge t)\kappa(t)$ with $\kappa(t) = I(C \ge T \mbox{ or } C \ge t)$ and
$e^{\Lambda_{a}^{*C}(t)}\kappa(t) = 1 - \int_0^t e^{\Lambda_{a}^{*C}(u)} \mathrm{d}M_a^{*C}(u)$, by iterated expectation, we have
\begin{align}
    \mathbb{P}\left\{\mathcal{S}_{a}(t; \widehat{\zeta}) - \mathcal{S}_{a}(t; \zeta^*)\right\} = \mathbb{P} & \left[ \frac{\delta}{\pi_{\delta}(X; \widehat{\eta})}\frac{A}{\pi_{A}(X;\widehat{\rho})} \left \{S_a(t,X) - S_{a}(t,X; \widehat{\beta_a})\right \} \right. \notag \\
           \ \  & + \tildelta d S_{a}(t,X; \widehat{\beta_a}) - S_a(t,X) \notag \\
            \ \ & - \frac{\delta}{\pi_{\delta}(X;\widehat{\eta})}\frac{A}{\pi_{A}(X;\widehat{\rho})}\int_0^t \frac{\mathrm{d}\widehat{M}_a^{C}(u;\widehat{\gamma_a})}{S_{a}^C(t,X;\widehat{\gamma_a})} \left\{S_a(t,X) -  \left. \frac{S_{a}(t,X;\widehat{\beta_a})}{S_{a}(u,X;\widehat{\beta_a})}\right\} \right] \notag \\
 \ = \ & \mathbb{P}\left [\left \{ \frac{\delta}{\pi_{\delta}(X;\widehat{\eta})}\frac{A}{\pi_{A}(X;\widehat{\rho})} - 1 \right\} \left \{S_a(t,X) - S_{a}(t,X; \widehat{\beta_a})\right \} \right ] \label{eq:B19} \\
             + \ & \mathbb{P}\left[ (\tildelta d - 1) S_{a}(t,X; \widehat{\beta_a})\right] \label{eq:B20}\\
             - \ & \mathbb{P}\left [ \frac{\delta}{\pi_{\delta}(X;\widehat{\eta})}\frac{A}{\pi_{A}(X;\widehat{\rho})}\int_0^t \frac{\mathrm{d}\widehat{M}_a^{C}(u;\widehat{\gamma_a})}{S_{a}^C(t,X;\widehat{\gamma_a})} \left\{\frac{S_a(t,X)}{S_a(u,X)} -  \frac{S_{a}(t,X;\widehat{\beta_a})}{S_{a}(u,X;\widehat{\beta_a})}\right\} \right ], \label{eq:B21} 
\end{align}
for $a \in \{0, 1\}$.
By Cauchy-Schwarz inequality and the positivity of $\pi_{\delta}(X)$ and $\pi_{A}(X)$, we have
\begin{align*}
    \eqref{eq:B19} & = \mathbb{P}\left [\left \{ \frac{\pi_{\delta}(X)\pi_{A}(X) - \pi_{\delta}(X;\widehat{\eta})\pi_{A}(X;\widehat{\rho})}{\pi_{\delta}(X;\widehat{\eta})\pi_{A}(X;\widehat{\rho})}\right\} \left \{S_a(t,X) - S_{a}(t,X; \widehat{\beta_a})\right \} \right ] \\
    & \lesssim ||\pi_{\delta}(X)\pi_{A}(X) - \pi_{\delta}(X;\widehat{\eta})\pi_{A}(X;\widehat{\rho})||\cdot||S_a(t,X) - S_{a}(t,X; \widehat{\beta_a}) || \\
    & \le \left \{ ||\pi_{\delta}(X) - \pi_{\delta}(X;\widehat{\eta})|| + ||\pi_{A}(X) - \pi_{A}(X;\widehat{\rho})||\right \}||S_a(t,X) - S_{a}(t,X; \widehat{\beta_a}) ||,
\end{align*}
where $\lesssim$ indicates that the inequality holds up to a multiplicative constant.
By iterated expectation, $\eqref{eq:B20} = 0$. Lastly, by Cauchy-Schwarz inequality and iterated expectation, we have
\begin{align*}
    \eqref{eq:B21} & \le \mathbb{P}\left [ \frac{\pi_{\delta}(X)}{\pi_{\delta}(X;\widehat{\eta})}\frac{\pi_A(X)}{\pi_{A}(X;\widehat{\rho})}\int_0^t \left|\left|\frac{\mathrm{d}\widehat{M}_a^{C}(u;\widehat{\gamma_a})}{S_{a}^C(t,X;\widehat{\gamma_a})}\right|\right|\cdot \left|\left|\frac{S_a(t,X)}{S_a(u,X)} -  \frac{S_{a}(t,X;\widehat{\beta_a})}{S_{a}(u,X;\widehat{\beta_a})}\right|\right| \right ] \\
    & \lesssim  \mathbb{P}\left [\int_0^t \left|\left|\mathrm{d}\widehat{M}_a^{C}(u;\widehat{\gamma_a})\right|\right|\cdot \left|\left|\frac{S_a(t,X)}{S_a(u,X)} -  \frac{S_{a}(t,X;\widehat{\beta_a})}{S_{a}(u,X;\widehat{\beta_a})}\right|\right| \right ],
\end{align*}
where the last inequality holds by the condition \textit{(C2)} in Theorem \ref{thm:efficiency bound}.
Therefore, $\mathbb{P} \left\{\mathcal{S}_{a}(t; \widehat{\zeta}) - \mathcal{S}_{a}(t; \zeta^*)\right\}$ is bounded above by $||\pi_{\delta}(X) - \pi_{\delta}(X;\widehat{\eta})||\cdot||S_a(t,X) - S_{a}(t,X; \widehat{\beta_a}) || + ||\pi_{A}(X) - \pi_{A}(X;\widehat{\rho})||\cdot||S_a(t,X) - S_{a}(t,X; \widehat{\beta_a})|| + \mathbb{P} \int_0^t ||\mathrm{d}\widehat{M}_a^{C}(u;\widehat{\gamma_a})||\cdot ||S_a(u,X)^{-1}S_a(t,X) -  S_{a}(u,X;\widehat{\beta_a})^{-1}S_{a}(t,X;\widehat{\beta_a})|| $.
It is negligible under the conditions \textit{(C1)} and \textit{(C3)} in Theorem \ref{thm:efficiency bound} thus $\mathbb{P}\left\{\vartheta_{\tau}(\widehat{\zeta}) - \vartheta_{\tau}(\zeta^*)\right \}$ is negligible as well and $\widehat{\theta}_{\tau}^{\text{ACW1}}$ achieves semiparametric efficiency. The proof for $\widehat{\theta}_{\tau}^{\text{ACW2}}$ is analogous by the small order difference between $\widehat{\theta}_{\tau}^{\text{ACW1}}$ and $\widehat{\theta}_{\tau}^{\text{ACW2}}$.

\end{document}

%% file: demo.tex
%% LyX 2.3.2 created this file.  For more info, see http://www.lyx.org/.
%% Do not edit unless you really know what you are doing.
\arrayrulecolor{white}
\begin{tikzpicture}[every node/.style=scale=1.1, node distance = 3.2cm, auto]
\tikzstyle{block} = [rectangle, draw, text width=18em, text centered, rounded corners, minimum height=3.5em]
\tikzstyle{every node} = [font = \footnotesize]
% Place nodes
\node [block] (A) {\begin{tabular}{c} 	% content
%	\textbf{Latent Data}\\
    \textbf{Target future patient population} \\
    	(Super-population)\\
	$\left\{T^1_i, T^0_i, X_i \right\}_{i=1}^\infty$
\end{tabular}
};
\node at (-4.5,-2) [block] (B) {	\begin{tabular}{c} 	% content
%	\textbf{Latent Data}\\
    \textbf{Finite RCT population}
	%$\left\{X_i,Y_i(0),Y_i(1)\right\}_{i=1}^\infty$
\end{tabular}
};
\node at (4.5,-2) [block] (C) {	\begin{tabular}{c} 	% content
%	\textbf{Latent Data}\\
    \textbf{Finite OS population}
	%$\left\{X_i,Y_i(0),Y_i(1)\right\}_{i=1}^\infty$
\end{tabular}
};
\node [block, below of=B] (D) {
	\begin{tabular}{l} 	% content
%	\textbf{Latent Data}\\
	$\left\{ T^1_i, T^0_i, X_i, \delta_i = 1,\widetilde{\delta}_i = 0 \right\}_{i=1}^n$
	\end{tabular}
};
\node  [block, below of = C] (E) {\begin{tabular}{l} 
%	\textbf{Latent Data}\\
	$\left\{T^1_i, T^0_i, X_i,  \delta_i =0,\widetilde{\delta}_i = 1\right\}_{i=n+1}^{n+m}$
\end{tabular}
};
\node at (-4.5,-3.6)(s-RCT){RCT Sampling: $\delta_i \sim$ Unknown};
\node at (4.5,-3.6)(s-RWE){OS Sampling: $\widetilde{\delta}_i\sim $ Known design};
\node at (-4.5,-6.6)(t-RCT) {RCT Treatment: $A_i \sim$ Randomization};
\node at (-4.5, - 6.95)(c-RCT) {Censoring:  noninformative $C$};
\node at (4.5,-6.8)(t-RWE) {OS Treatment: $A_i \sim$ Unknown};
%\node at (-2.5,-5){$A_i \sim$ Randomized};
%\node at (6.5,-5){$A_i \sim$ Unknown};
\node [block, below of=D] (F) {
	\begin{tabular}{c}
	\textbf{Observed RCT Sample}\\
	$\left\{U_i, \Delta_i, A_i, X_i, \delta_i = 1,\widetilde{\delta}_i = 0\right\}_{i=1}^n$
	\end{tabular}
};
\node [block, below of=E] (G) {
\begin{tabular}{c}
 \textbf{Observed OS Sample}\\
$\left\{X_i, \delta_i = 0,\widetilde{\delta}_i=1\right\}_{i=n+1}^{n+m}$
\end{tabular}
};
% Draw edges
\draw [->,shorten >=4pt] 

(A) edge (B) 
(A) edge (C)
(B) edge (s-RCT)
(s-RCT) edge (D)
(C) edge (s-RWE)
(s-RWE) edge (E)
(D) edge (t-RCT)
(E) edge (t-RWE)
(c-RCT) edge (F)
(t-RWE) edge (G); 
\end{tikzpicture}